\documentclass[journal=jcisd8,manuscript=article]{achemso}

\usepackage{graphicx}
\usepackage{amsmath,amssymb}
\usepackage{xcolor}
\usepackage{booktabs}
\usepackage{float}
\usepackage{url}
\usepackage{xspace}
\newcommand{\sss}{ShEPhERD-score\xspace}
\renewcommand{\mit}{Massachusetts Institute of Technology, Cambridge, Massachusetts 02139, United States}

\newcommand{\chem}{Department of Chemistry, \mit}
\newcommand{\cheme}{Department of Chemical Engineering, \mit}

\author{Nate Somers}
\altaffiliation{These authors contributed equally to this work.}
\affiliation{\chem}

\author{Kento A. Abeywardane}
\altaffiliation{These authors contributed equally to this work.}
\affiliation{\cheme}

\author{Connor W. Coley}
\affiliation{\cheme}
\altaffiliation{Department of Electrical Engineering and Computer Science, \mit}

\author{Masha Elkin}
\email{melkin@mit.edu}
\affiliation{\chem}

\title{ShEPhERD-score: Fast, adaptable three-dimensional alignment of molecular features}

\begin{document}


\begin{center}
\includegraphics[width=0.78\linewidth]{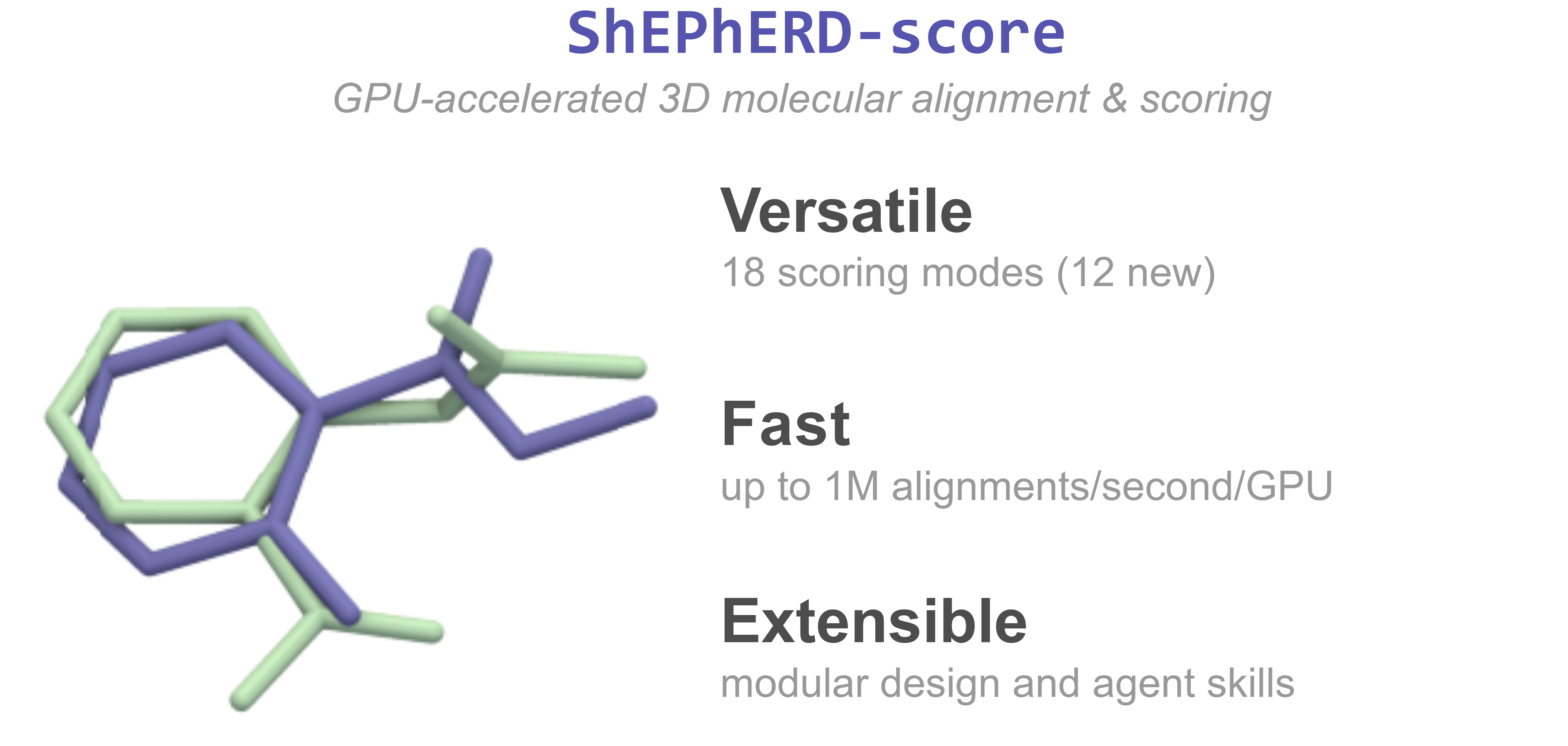}
\end{center}

\begin{abstract}
\noindent

Three-dimensional (3D) molecular similarity scoring and alignment are widely employed in ligand-based virtual screening and molecular design, and are increasingly used for training and evaluating machine learning models. Existing open-source methods utilize various molecular representations spanning shape, electrostatics, pharmacophores, and hybrid representations, but their implementations as independent software packages with distinct application programming interfaces (APIs), optimizers, and limited GPU support make them difficult to compare, combine, or extend. Here, we develop \sss{} into a unified, modular platform for differentiable 3D molecular similarity scoring and alignment by combining diverse molecular representations with high-performance Numba CPU and Triton GPU kernels to support both large-scale library screening and pairwise similarity workflows. In matched comparisons with open-source 3D similarity tools on the DUDE-Z virtual screening benchmark, \sss{} performed comparably or better in eight of nine cases. Across the scoring modes in \sss{}, throughput ranges from 17{,}800 to 1.03 million alignments/second for ten-million-conformer libraries using a single NVIDIA L40S GPU. Moreover, we found that the most effective scoring mode varied by protein target, indicating that target-specific calibration against validation data may be valuable before large-scale screening and underscoring the need for a unified platform that provides diverse molecular representations. Together, these capabilities establish \sss{} as a single, extensible toolkit for rapid virtual screening and machine learning workflows.

\end{abstract}

\section{Introduction}
\label{sec:intro}

The three-dimensional (3D) structures of molecules define the spatial arrangement of steric and electronic features, which governs the intermolecular interactions and chemical reactivity that underlie their function. For example, proteins recognize ligands through their shape and the placement of polar and hydrophobic groups via non-covalent interactions.\cite{bissantzMedicinalChemistsGuide2010} Comparing these features through 3D molecular alignment and similarity scoring is therefore a fundamental operation in cheminformatics. This process superimposes two 3D conformations, quantifying similarity by calculating differences in molecular shape, electrostatic potential (ESP), pharmacophores, or combinations thereof.\cite{goodUtilizationGaussianFunctions1992, nicholls2010shape, shin2015threedimensional, leachThreeDimensionalPharmacophoreMethods2010, jiangComprehensiveComparativeAssessment2021,lo3DChemicalSimilarity2016} The alignment itself is an optimization problem in which the relative orientation of the two conformations is varied to maximize their similarity. These similarity measures are routinely used to prioritize compounds during virtual screening,\cite{rush2005scaffold, fiedlerMAP4K4InhibitionPromotes2019, petrovicVirtualScreeningCloud2022, langdonScaffoldFocusedVirtualScreening2013, bostromPotentFibrinolysisInhibitor2013, hurtleLigandBasedVirtualScreening2024} identify bioisosteric replacements in medicinal chemistry,\cite{cheeseright2006field, schuffenhauerSimilaritySearchingFiles1999, jacksonApplicationMolecularModelingScaffoldHopping2022} and analyze structure-activity relationships to provide 3D insight to chemists.\cite{medinafranco2009consensus, clevesChemicalProteinStructural2015, correyExplorationStructureactivityRelationships2025} Increasingly, these alignment scores have been used as objectives for machine-learned screening methods\cite{winterLearningCompressedShapeAware2026, lzicarCHEESE3DShape2024} or evaluation metrics for molecular generative models.\cite{adams2025shepherd,Abeywardane2026shepherd2,irwin2026ensemble}

Existing 3D molecular alignment software supports molecular representations based on molecular shape,\cite{grantGaussianDescriptionMolecular1995, grant1995gaussian, openeye_fastrocs, jainMorphologicalSimilarity3D2000, haque2010paper,atwi2024roshambo} ESP,\cite{bolcato2022espsim,cleves2019esim} pharmacophores,\cite{spragueAutomatedChemicalHypothesis1995, taminau2008pharao,wahl2024phesa,dixon2006phase,koes2011pharmer} or hybrid scoring functions.\cite{liu2011shafts,vainio2009shaep,armstrong2010electroshape,tosco2011o3a,yan2014gwega,douguet2020sensaas,grisoniScaffoldHoppingNatural2018} Among these, OpenEye ROCS and FastROCS are commercial tools widely used for rapid 3D molecular similarity screening with volumetric representations of shape and directionless pharmacophores.\cite{openeye_fastrocs} Open-source tools, including ROSHAMBO2,\cite{atwi2024roshambo, roshambo2_2025} ShaEP,\cite{vainio2009shaep} PheSA,\cite{wahl2024phesa} Pharao/Align-it,\cite{taminau2008pharao} RDKit (\texttt{rdShapeAlign}),\cite{rdkit2026} and ESP-Sim,\cite{bolcato2022espsim} each support one or two of these representations, and only ROSHAMBO2 runs on GPUs. However, these tools have largely been developed independently;  each similarity metric is implemented with its own alignment algorithm in its own software package, often with distinct APIs, optimization strategies, and hardware support. As a result, evaluating multiple definitions of molecular similarity---or developing new ones---remains difficult, particularly when methods cannot be compared under consistent alignment and computational settings. This limitation matters because no single 3D similarity tool performs best for every target class or evaluation metric,\cite{jiangComprehensiveComparativeAssessment2021} motivating the need for flexible frameworks that support diverse molecular representations.

Rather than treating each similarity measure as a distinct algorithm, we formulate the alignment of molecular conformations as the optimization of arbitrary differentiable similarity objectives over rigid body translations and rotations. This abstraction separates the optimization procedure from the scoring function itself, allowing new objectives to be implemented by defining only a differentiable scoring function while reusing the same optimization framework. The resulting implementation naturally supports automatic differentiation, efficient batched execution on CPUs and GPUs, and straightforward composition of multiple scoring terms. This design, combined with code-generation agent skills, allows rapid prototyping, validation, and optimization of new similarity objectives with minimal manual implementation effort.\cite{chen2021codex, swe-agent, chemcrow} 

Herein we describe \sss{} as a robust, open-source implementation for differentiable 3D molecular similarity scoring and alignment. This work builds on ShEPhERD and ShEPhERD--2, our diffusion models for interaction-conditioned 3D molecule generation.\cite{adams2025shepherd,Abeywardane2026shepherd2} While six scoring functions were released alongside these generative models, they were not optimized for speed or modularity. In this work, we expand those capabilities for high-throughput virtual screening by introducing twelve new scoring modes and accelerating their computation. The eighteen modes evaluated here include shape, ESP, pharmacophore, and strategic combinations thereof via composite scoring functions, all through a common Python interface. We demonstrate that this extensible framework preserves competitive virtual screening performance while achieving substantial throughput improvements through optimized Triton GPU and Numba CPU kernels, reaching 17{,}800 to 1.03 million alignments/second for ten-million-conformer libraries on a single NVIDIA L40S GPU. The same representations can also compare 3D properties beyond those relevant to protein--ligand recognition, such as condensed Fukui reactivity indices and Crippen molar-refractivity contributions.

\section{Alignment Modes and Workflow}
\label{sec:modes}

\begin{figure}[H]
  \centering
  \includegraphics[width=\textwidth]{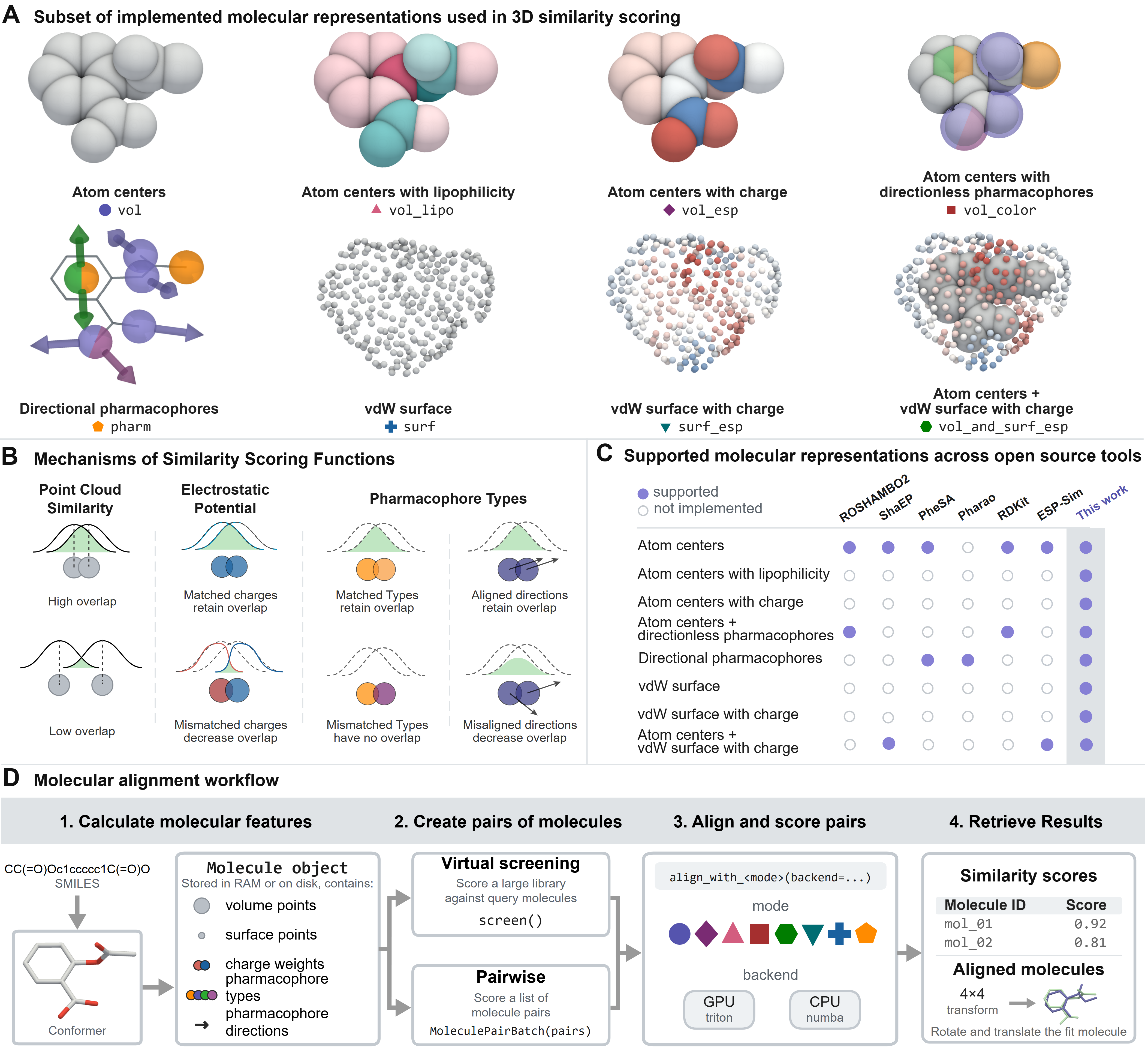}
  \caption{\textbf{Overview of \sss{}}.
\textbf{(A)} Representative subset of supported molecular representations.
  \textbf{(B)} Scoring functions measure Gaussian pairwise overlap with weighting that rewards matching charges, assigned type, or vector directionality.
  \textbf{(C)} \sss{} supports broad coverage of 3D similarity scoring functions compared to existing open-source tools. 
  \textbf{(D)} Workflow for molecular alignment and similarity scoring.
  }
  \label{fig:modes}
\end{figure}

The central abstraction in \sss{} is the separation of molecular representation, similarity scoring, and rigid-body optimization. Molecular alignment is formulated as the maximization of a differentiable similarity objective over SE(3),\cite{grant1995gaussian, adams2025shepherd} allowing different representations and scoring functions to use a common optimization framework. The modes described here primarily represent molecules as 3D point clouds associated with isotropic Gaussian functions\cite{grantGaussianDescriptionMolecular1995, grant1995gaussian} and, where applicable, chemical features or directional information.\cite{wahl2024phesa, adams2025shepherd} Figure~\ref{fig:modes}A summarizes the representations used by a representative set of eight of the eighteen modes evaluated in this work. Similarity scores are constructed from Gaussian overlaps with mode-specific weighting and normalization.\cite{vainio2009shaep, wahl2024phesa} Each mode is therefore defined by its point representation and its weighting function.

The basis of each molecular representation is defined by a 3D point cloud $A \in \mathbb{R}^{n_A \times 3}$ where each point $\mathbf{r}_i$ is an isotropic Gaussian in $\mathbb{R}^3$.\cite{goodUtilizationGaussianFunctions1992, grant1995gaussian} The first-order Gaussian overlap between two point clouds $A$ and $B$ is
\begin{equation}
    O_{AB} =
    \sum_{a \in A}\sum_{b \in B}
    w_{ab}
    \frac{\pi^{3/2}}{(2\alpha)^{3/2}}
    \exp\left(
    -\frac{\alpha}{2}
    \left\lVert \mathbf{r}_a-\mathbf{r}_b \right\rVert^2
    \right),
\end{equation}
where $\alpha$ is the Gaussian width associated with the point representation and $w_{ab}$ is a weighting function that defines the contribution of each point pair (Figure~\ref{fig:modes}B). This weight rewards point pairs with matching charges, matching feature types, or aligned feature vectors. In most cases, the overlap is normalized using the symmetric Tanimoto coefficient,
\begin{equation}
    S(A,B)=
    \frac{O_{AB}}
    {O_{AA}+O_{BB}-O_{AB}},
\end{equation}
which yields values between zero and one and defines the similarity score. Composite scoring functions combine separately normalized terms. Because molecular similarity depends on the relative orientation of the two molecules, alignment corresponds to finding the rigid-body transformation that maximizes this similarity,
\begin{equation}
    S^*(A,B)=
    \max_{R,t}\, S(RA+t,B),
\end{equation}
where $R \in SO(3)$ and $t \in \mathbb{R}^3$ define the rotation and translation applied to molecule $A$.

Using unit weights ($w_{a,b}=1$) on heavy-atom centers yields volumetric shape similarity (\texttt{vol}), while applying the same formulation to points sampled on the solvent-accessible surface scores surface similarity (\texttt{surf}). Charge or electrostatic similarity is incorporated by weighting Gaussian overlaps by the difference in charge $\mathbf{c} \in \mathbb R^{n_A}$ at each of those locations ($w_{a,b} = \exp\left(-\|c_a - c_b\|^2 / \lambda \right)$), giving \texttt{vol\_esp} and \texttt{surf\_esp}, while \texttt{vol\_and\_surf\_esp} combines volumetric shape with surface electrostatics in a composite score following the ShaEP formulation.\cite{vainio2009shaep} Pharmacophore similarity (\texttt{pharm}) represents molecules using a set of feature types $\mathcal{M}$, including directional features (e.g., hydrogen-bond donors, acceptors, and aromatic rings) and non-directional features (e.g., hydrophobic and ionic groups). Similarity is computed only between compatible feature types leading to a slightly different similarity aggregation scheme:
\begin{equation}
    S_{\texttt{pharm}}(A,B)=
    \frac{ \sum_{m \in \mathcal{M}} V_{AB;m}}
    {\sum_{m \in \mathcal{M}} \left(V_{AA;m}+V_{BB;m}-V_{AB;m}\right)}.
\end{equation}
An additional direction-dependent weighting is applied to features with associated vector directions $\mathbf{V_A} = \{\mathbf{v_1}, \dots, \mathbf{v_a}\}$ where $\mathbf{v_a} \in \mathbb{S}^{2}$ and $w_{a,b;m} = \frac{\mathbf{v_{a,m}}^{\top}\mathbf{v_{b,m}} + 2}{3}$.\cite{wahl2024phesa} For aromatic ring features, we use the absolute value of the dot product $\mathbf{v_{a,m}}^{\top}\mathbf{v_{b,m}}$, because the sign of a ring's normal vector is arbitrary and an aromatic ring can interact through either face.

This framework enables new similarity objectives to be implemented with minimal additional infrastructure. \texttt{vol\_color} extends volumetric shape similarity by including directionless pharmacophoric features in a ROCS-color-like formulation,\cite{hawkins2007rocs} while \texttt{vol\_lipo} weights atom-centered Gaussian overlaps by Crippen lipophilicity parameters\cite{wildman1999crippen} to favor alignments between corresponding hydrophobic and polar regions. By introducing these formulations, \sss{} extends the combined representation coverage of existing open-source tools, doubling the number of supported representations (Figure~\ref{fig:modes}C).

These scoring functions share a common SE(3) optimization, batching, and hardware acceleration infrastructure. Molecule orientations are parameterized by a unit quaternion and translation vector, and multiple initial orientations are refined in parallel using the Adam optimizer\cite{kingma2015adam} to identify the highest-scoring alignment. The workflow begins by converting molecules into reusable molecular profiles that cache the features required for alignment (Figure~\ref{fig:modes}D). These profiles can then be aligned either in-memory using \texttt{MoleculePairBatch} for dense pairwise comparisons or streamed directly from disk using \texttt{ProfileStore} for large-scale virtual screening.

To enable users to quickly prototype and optimize new scoring modes, we include structured instructions that allow coding agents to implement them from a natural language prompt. New objectives are first expressed as differentiable PyTorch reference implementations. Optimized Numba and Triton kernels are then generated and automatically validated against the reference implementation. Twelve new scoring modes including \texttt{vol\_color}, \texttt{vol\_fukui}, and \texttt{pharm\_tversky} were developed using this workflow (Supporting Information, Section~\ref{sec:si-modes}).

\section{Benchmarking Similarity Objectives}
\label{sec:enrichment}

We assessed the accuracy of \sss{} scoring functions on the DUDE-Z benchmark of 43 protein targets with associated active ligands and property-matched decoys.\cite{stein2021dudez} This benchmark measures how well a scoring function discriminates known actives from decoys given a single known active, and its common use for evaluating 3D ligand similarity scoring functions provides a point of comparison.\cite{atwi2024roshambo, roshambo2_2025, shmilovich2026struct2query} Although property-matched decoys can favor methods that represent molecular connectivity,\cite{sieg2019bias, chen2019hiddenbias} this well-established benchmark enables a fair comparison among methods evaluated on the same query ligand and library conformers. We compared our scoring functions directly against open-source alignment packages which use similar molecular representations and scoring functions to test whether the accelerated implementation retains their retrieval accuracy.

\begin{figure}[H]
  \centering
  \includegraphics[width=\textwidth]{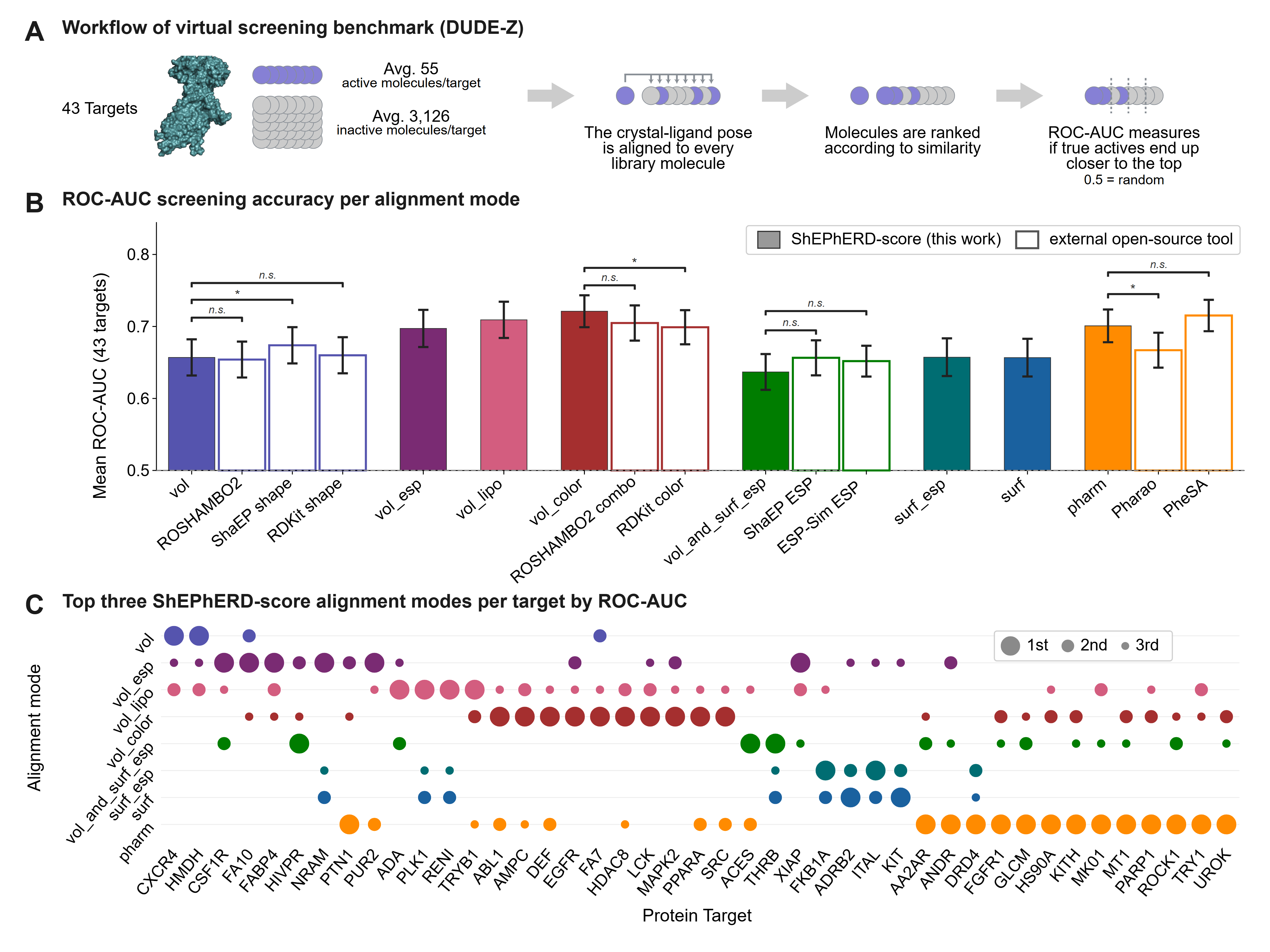}
  \caption{\textbf{Virtual screening benchmark on DUDE-Z.}
  \textbf{(A)} For each of 43 targets, the co-crystallized ligand pose is aligned against up to 50 conformers of every library compound, and molecules are ranked by 3D similarity.
  \textbf{(B)} Mean ROC-AUC across the 43 targets for each \sss{} mode (solid) and corresponding open-source implementations (outlined): ROSHAMBO2,\cite{roshambo2_2025} ShaEP,\cite{vainio2009shaep} Pharao/Align-it,\cite{taminau2008pharao} PheSA,\cite{wahl2024phesa} RDKit,\cite{rdkit2026} and ESP-Sim.\cite{bolcato2022espsim} \texttt{surf}, \texttt{surf\_esp}, \texttt{vol\_esp}, and \texttt{vol\_lipo} have no direct counterpart among these tools. Brackets are paired $t$-tests ($^{*}p<0.05$, n.s.\ $=$ not significant); error bars are the standard error of the mean across targets.
  \textbf{(C)} Per-target top-three modes by ROC-AUC where circle size denotes relative rank.}
  \label{fig:enrichment}
\end{figure}

Following an established DUDE-Z evaluation protocol, we used, for each of the 43 protein targets, the pose of a ligand co-crystallized with that target as the query (Figure~\ref{fig:enrichment}A).\cite{atwi2024roshambo, roshambo2_2025, shmilovich2026struct2query} The query was aligned against up to 50 generated conformers per library ligand to account for molecular flexibility, with the maximum similarity score retained for ranking. Retrieval performance was evaluated using the area under the receiver operating characteristic curve (ROC-AUC), which measures the probability that a randomly selected active is ranked above a randomly selected decoy.\cite{jain2008recommendations} All scoring functions used the same conformers to isolate differences in molecular representation, scoring, and search effort. Alignment settings for enrichment and throughput measurements are reported in the Supporting Information (Sections~\ref{sec:si-enrichment-protocol} and \ref{sec:si-protocol}).

In this benchmark, the \sss{} alignment modes achieved significantly higher ROC-AUC in two of nine comparisons (Figure~\ref{fig:enrichment}B), comparable performance in six, and lower ROC-AUC in one. Specifically, \texttt{vol\_color} outperformed RDKit color and \texttt{pharm} outperformed Pharao, whereas ShaEP shape outperformed \texttt{vol} (all $p<0.05$, two-sided paired $t$-test across 43 targets). The inclusion of hydrogen atoms in ShaEP may contribute to its higher ROC-AUC, whereas \texttt{vol} uses heavy atoms only. The accelerated implementation therefore retains the retrieval accuracy of existing open-source tools in eight of nine comparisons, and introduces new alignment modes which lack existing comparators.

Virtual screening aims to prioritize actives at the top of a ranked library for further downselection and, ultimately, synthesis and experimental testing. ROC-AUC assesses overall ranking accuracy, whereas early enrichment measures how many actives are recovered among the top-ranked compounds. We therefore also report the 1\% enrichment factor (EF1\%) and hit rate, which normalizes EF1\% by its maximum attainable value for that library (Supporting Information, Section~\ref{sec:si-enrichment}).\cite{jain2008recommendations, jiangComprehensiveComparativeAssessment2021} EF1\% and hit rates for every method are in Table~\ref{tab:si-modes}, with per-target EF1\% in Tables~\ref{tab:si-ef1-main} and \ref{tab:si-ef1-si}. On EF1\%, \sss{} achieved comparable or higher enrichment than existing methods in all nine matched comparisons (Figure~\ref{fig:si-ef1}).

To investigate how retrieval depends on the molecular representation, we compared the \sss{} modes on each target and found that the best-performing mode varied (Figure~\ref{fig:enrichment}C). Among the \sss{} modes shown in Figure~\ref{fig:enrichment}B, weighted volumetric representations gave the strongest average benchmark performance. \texttt{vol\_color} achieved the highest mean ROC-AUC of 0.721, compared with 0.657 for volumetric shape alone, indicating that adding molecular attributes can improve retrieval. However, the benefit depended on both the attributes used and the target. Each of the eight modes ranked first for at least one target and second for at least one target, demonstrating that no single scoring function was optimal across all targets. Where target-specific validation data are available, the scoring mode and any term weights can therefore be selected as hyperparameters and confirmed on an independent test set before prospective use. Because every mode is called through the same interface, this selection requires only changing the mode argument.

\section{Throughput and Scaling}\label{sec:speed}

Enumerated make-on-demand catalogs of molecules now hold tens of billions of readily synthesizable compounds,\cite{lyu2019ultralarge, grygorenko2020enamine, tingle2023zinc22} and the combinatorial spaces that generate them imply hundreds of trillions or more,\cite{hoffmann2019chemicalspace, warr2022ultralarge, klein2026chemicalspaces} making throughput a practical consideration for 3D similarity screening. To support large-scale use, we implemented score and gradient evaluation in Numba CPU\cite{lam2015numba} and Triton GPU kernels, while grouping conformer pairs by size to improve GPU utilization.\cite{tillet2019triton}

\begin{figure}[H]
  \centering
  \includegraphics[width=\textwidth]{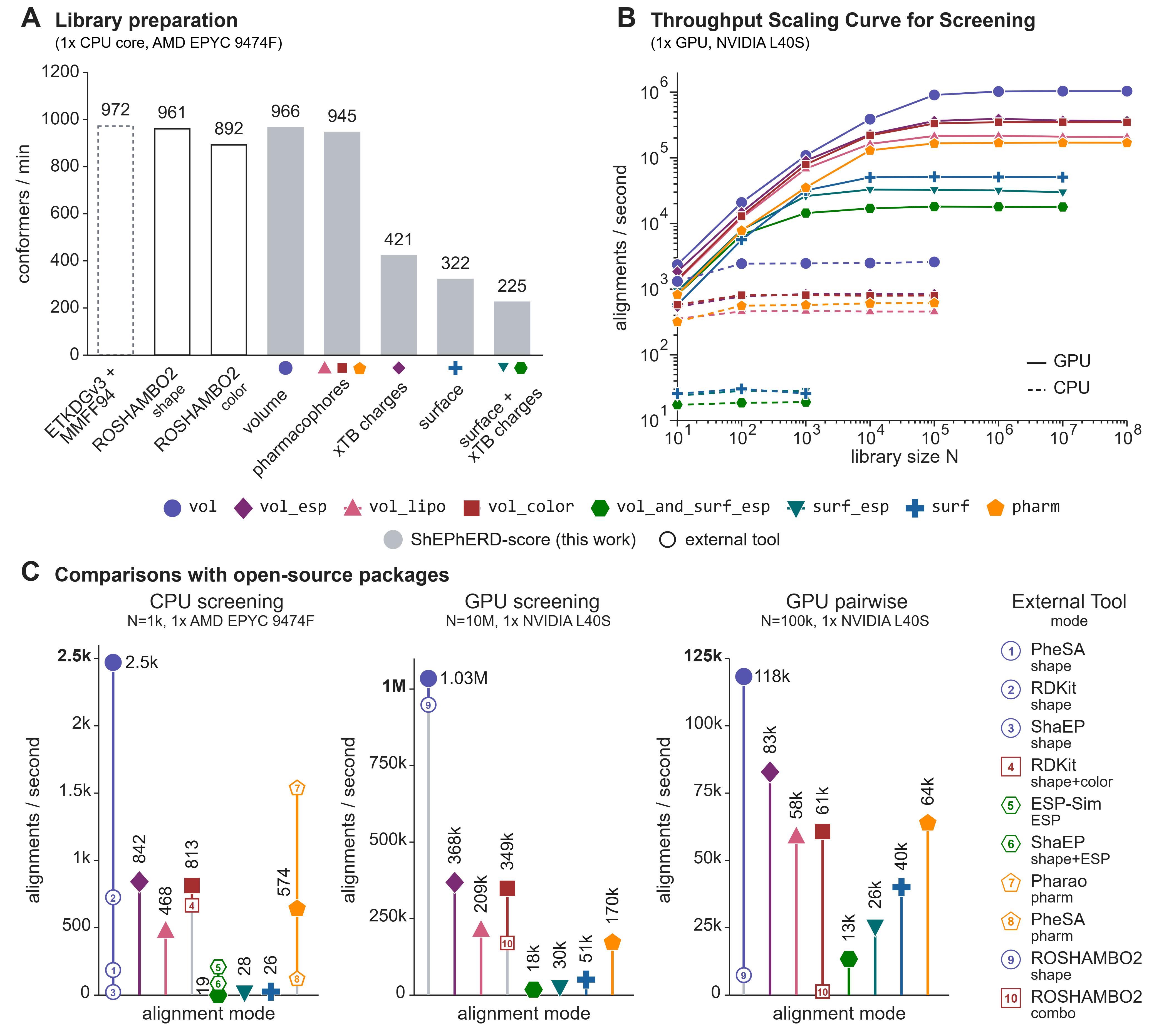}
  \caption{\textbf{\sss{} preparation and alignment throughput.}
  \textbf{(A)} Preparation throughput on one CPU core for 100 compounds with 50 conformers each. All rates include shared ETKDGv3 embedding, MMFF94 relaxation, and aromaticity restoration (dashed outline), plus any additional preparation. Symbols indicate the similarity calculation mode. Feature extraction by external tools which use structure data file (SDF) inputs is excluded from preparation timings because it occurs during alignment.
  \textbf{(B)} Screening throughput versus library size on a GPU (solid) and one CPU core (dashed).
  \textbf{(C)} Throughput comparisons for single-core CPU screening ($N=10^{3}$), GPU screening ($N=10^{7}$), and GPU pairwise alignment ($N=10^{5}$). Measured on an AMD EPYC 9474F CPU and an NVIDIA L40S GPU.}
  \label{fig:speed}
\end{figure}

Screening large libraries requires the initial featurization of all candidate ligands and their conformers. However, because different open-source packages perform this step at different stages of the alignment workflow, we compare the time required to compute and store molecular objects. Figure~\ref{fig:speed}A reports preparation throughput in conformers per minute on one CPU core, measured over 100 compounds with 50 conformers each. All packages compared here share a baseline preparation time of 61.7\,ms per conformer for ETKDGv3 embedding and MMFF94 relaxation. \sss{} volumetric shape and pharmacophore feature extraction add minimal overhead to open-source package preparation times at 62.1 and 63.5\,ms per conformer, respectively. However, modes requiring GFN2-xTB charges, surfaces, or both take an average of 142.4, 186.6, and 267.1\,ms per conformer, respectively. Preparation is therefore a one-time cost comparable to that of existing tools, except for the charge- and surface-based modes.

\sss{} provides two alignment workflows. The \texttt{screen} workflow aligns a single query conformer against a large precomputed library to find compounds similar to a known active. Molecular profiles are stored on disk and streamed into memory during screening, so a library is featurized once and searched repeatedly. The \texttt{pairwise} workflow holds two sets of conformers in memory and aligns specified pairs or batches, processing comparisons larger than host memory as separate blocks. The \texttt{pairwise} workflow is useful for smaller library comparisons, cheminformatics analysis,\cite{jainMorphologicalSimilarity3D2000} and machine-learning applications that require large numbers of molecular similarity evaluations.\cite{bolcato2022espsim, lzicarCHEESE3DShape2024, winterLearningCompressedShapeAware2026} Both workflows contain accelerated implementations on GPU, to maximize throughput, and on CPU, to maximize accessibility. 

We first measured how screening throughput scales with library size on one GPU and one CPU core (Figure~\ref{fig:speed}B). GPU screening throughput initially increases with library size as larger batches improve GPU utilization, then approximately plateaus above $N=10^{5}$ and remains high through $N=10^{7}$ conformers for all eight modes, whereas single-core CPU throughput changes less than twofold across library sizes. At $N=10^{7}$, GPU screening was 296 to 1{,}972 times faster than single-core CPU screening at $N=10^{3}$, depending on the mode. Screening throughput can be increased further by distributing the library across additional CPU workers or GPUs. For \texttt{vol}, throughput increased 54.7-fold with 64 CPU workers and 3.9-fold with four GPUs relative to the corresponding single-worker runs (Figure~\ref{fig:si-multiproc}).

Compared to open-source tools, \sss{} is typically faster on a single AMD EPYC 9474F CPU core (Figure~\ref{fig:speed}C, left). Volumetric shape alignment reaches 2{,}470 alignments/s, compared with 726 for RDKit,\cite{rdkit2026} 186 for PheSA, and 22.5 for ShaEP; shape-plus-color alignment reaches 813 alignments/s, compared with 667 for RDKit; and pharmacophore alignment reaches 574 alignments/s, compared with 180 for PheSA. ShaEP and ESP-Sim are faster for combined volume-and-surface electrostatic similarity (22.5 and 187 versus 19 alignments/s), although ESP-Sim scores electrostatic similarity after Crippen O3A alignment and therefore uses a different alignment objective. Pharao/Align-it is faster for pharmacophore alignment (1{,}533 alignments/s), superposing matched features by their centroids and optimizing rotation only, but has significantly lower DUDE-Z ROC-AUC than \texttt{pharm} ($p=0.018$). Single-core CPU screening with \sss{} is therefore competitive with existing CPU tools.

Existing open-source GPU implementations support only volumetric shape and shape-plus-color similarity, so direct comparisons used these two modes (Figure~\ref{fig:speed}C, middle). On a single NVIDIA L40S GPU screening $N=10^{7}$ conformers, \sss{} reaches 1{,}034{,}608 alignments/s for volumetric shape, comparable to the 948{,}736 alignments/s of ROSHAMBO2, and 348{,}793 alignments/s for shape-plus-color similarity, twice the 170{,}289 alignments/s of ROSHAMBO2. For the other six modes in Figure~\ref{fig:speed} and the ten additional modes (Table~\ref{tab:si-throughput}), \sss{} provides GPU screening where no open-source GPU implementation is available.

The advantage of \sss{} over existing open-source software is especially evident for exhaustive pairwise GPU alignment. In this setting, \sss{} processes pairs from multiple reference queries simultaneously, while ROSHAMBO2 processes each query separately. On the same 100{,}000 ordered conformer pairs, \sss{} was 16.0$\times$ faster for \texttt{vol} (118{,}293 versus 7{,}415 alignments/s) and 45.9$\times$ faster for \texttt{vol\_color} (60{,}728 versus 1{,}323 alignments/s) than the corresponding ROSHAMBO2 scoring functions (Figure~\ref{fig:speed}C, right). At these rates, one GPU computes 3D similarities for one million conformer pairs in 8.5 to 75\,s across the eight modes, which makes pairwise scoring practical for building machine-learning training sets and evaluating generative models.

\section{Conclusions}
\label{sec:conclusions}

This work represents a substantially expanded and improved software package for 3D molecular alignment that is well equipped for modern high-throughput workflows. By formulating 3D molecular alignment as the optimization of modular, differentiable objectives over $\text{SE}(3)$, \sss{} decouples the definition of 3D chemical similarity from the underlying optimization, batching, and hardware-acceleration infrastructure. This formulation enabled speedups up to 46-fold relative to best-in-class open-source tools while introducing twelve new scoring modes and several new representations. Benchmarking on DUDE-Z, we achieve comparable or better performance than the corresponding open-source tools in eight of nine comparisons. Because we revealed that no single representation is universally optimal across protein targets, the ability to seamlessly switch, combine, or prototype scoring functions within one high-throughput framework offers a practical advantage for ligand-based virtual screening and is synergistic with 3D generative modeling workflows. Future efforts will expand the integration of this work with generative models and synthetic planning to enable molecular optimization. Ultimately, \sss{} provides an extensible, open-source toolkit that bridges foundational 3D cheminformatics and modern machine-learning pipelines for molecular design.

\section*{Data and Software Availability}
\sss{} is released in the open-source \texttt{shepherd-score} v2.0 package under the MIT license at \url{https://github.com/coleygroup/shepherd-score}. The benchmark harnesses, figure scripts, and results are available at \url{https://github.com/ElkinGroup/Shepherd-Score-Benchmarks}. Software versions and commits are listed in Table~\ref{tab:si-software} in the Supporting Information.

\begin{acknowledgement}
C.W.C.\ thanks the Machine Learning for Pharmaceutical Discovery and Synthesis (MLPDS) consortium for its support. N.S. is supported as an XtalPi AI for Science Fellow. K.A.A.\ received funding from the MIT Schwarzman College of Computing as a Future Research Cohort Fellow. We thank Patrick Riley for contributions to the ShEPhERD-score package, development of the \texttt{vol\_avoid} scoring function, and insightful discussions. We also thank Kenji Walker for helpful discussions of scoring functions and contributions to profile extraction.
\end{acknowledgement}

\begin{suppinfo}
Additional alignment modes, benchmark preparation and alignment protocols, per-target enrichment results, and throughput and scaling measurements (PDF)
\end{suppinfo}

\begin{tocentry}
\centering
\includegraphics[width=3.25in,height=1.75in,keepaspectratio]{ACS_methylsalicylate_stick.png}
\end{tocentry}

\bibliography{references}

%
\clearpage
\setcounter{figure}{0}\renewcommand{\thefigure}{S\arabic{figure}}
\setcounter{table}{0}\renewcommand{\thetable}{S\arabic{table}}
\newsavebox{\sitablebox}
\newcommand{\fitsitable}[1]{%
  \sbox{\sitablebox}{#1}%
  \ifdim\wd\sitablebox>\linewidth
    \sbox{\sitablebox}{\resizebox{\linewidth}{!}{\usebox{\sitablebox}}}%
  \fi
  \ifdim\dimexpr\ht\sitablebox+\dp\sitablebox\relax>0.72\textheight
    \resizebox*{!}{0.72\textheight}{\usebox{\sitablebox}}%
  \else
    \usebox{\sitablebox}%
  \fi
}
\section{Supporting Information}
\label{sec:si}
\makeatletter
\let\@startsection\acs@startsection
\makeatother
\setcounter{secnumdepth}{3}
\setcounter{subsection}{0}
\renewcommand{\thesubsection}{S\arabic{subsection}}
\renewcommand{\thesubsubsection}{\thesubsection.\arabic{subsubsection}}

Section~\ref{sec:si-modes} describes the agent skills and the ten additional alignment modes not discussed in detail in the main text. Section~\ref{sec:si-enrichment} presents virtual-screening enrichment methods and results. Section~\ref{sec:si-throughput} presents alignment throughput methods, results for the additional modes, and multi-process/device screening.

\subsection{Additional alignment modes}
\label{sec:si-modes}

\sss{} separates mode-specific score and gradient kernels from the shared optimization and batching infrastructure. We provide two packaged agent skills to facilitate the development of new scoring functions from natural language. \texttt{design-scoring-mode} implements a scoring function with autograd via PyTorch or JAX. \texttt{accelerate-scoring-mode} implements a Numba or Triton backend for hardware acceleration via analytic gradients. Optimized score and gradient implementations are checked against the differentiable reference at fixed poses before use in alignment. Using this workflow, \texttt{vol\_color}, \texttt{vol\_lipo}, and ten extra modes (described below) were added to \sss{}. 

Four of the ten additional modes combine volumetric shape with additional molecular features. \texttt{vol\_pharm} combines shape with directional pharmacophore similarity. In these benchmarks, it uses the same pharmacophore feature definitions as the \texttt{vol\_color} mode, but includes directional weighting that \texttt{vol\_color} omits. \texttt{vol\_atomtype} decorates volumetric shape with discrete element types while \texttt{vol\_mr} uses per-atom Crippen molar-refractivity contributions. \texttt{vol\_fukui} uses a condensed Fukui reactivity field $\Delta f = f^{+}-f^{-}$ where $f^{+}=q_{N}-q_{N+1}$ and $f^{-}=q_{N-1}-q_{N}$ are the condensed Fukui functions obtained from three GFN2-xTB single points for a given conformer (the formal molecular charge and the adjacent charge states). The descriptor is positive at electrophilic centers and negative at nucleophilic centers.

The other six modes implement asymmetric Tversky similarity $T_v = O_{AB}/(\alpha\,O_{AA}+(1-\alpha)\,O_{BB})$ where $\alpha=0.95$ for \texttt{surf\_tversky}, \texttt{surf\_esp\_tversky}, \texttt{vol\_and\_surf\_esp\_tversky}, \texttt{vol\_color\_tversky}, \texttt{vol\_lipo\_tversky}, and \texttt{pharm\_tversky} instead of Tanimoto similarity. This was done because Tversky similarity rewards query molecules that overlap well with the reference, but receive a smaller penalty for ``extra'' volume or features that extend beyond the reference, which is useful for retrieving larger molecules that contain the reference scaffold. The \texttt{vol} and \texttt{vol\_esp} modes already included Tversky normalization and we extend it to the six modes listed above. In \texttt{vol\_and\_surf\_esp\_tversky} the Tversky reduction applies to the shape channel only, because the ShaEP-style surface-potential agreement channel is a masked point-to-point average rather than an overlap ratio.

\subsection{Virtual-screening enrichment}
\label{sec:si-enrichment}

\subsubsection{Benchmark protocol and metrics}
\label{sec:si-enrichment-protocol}

We apply a uniform benchmarking protocol for all software across the 43 DUDE-Z targets, using each target's co-crystallized ligand pose as the reference. The actives and property-matched decoys were downloaded from the DUDE-Z website (\url{https://dudez.docking.org/}) on 29 September 2026,\cite{stein2021dudez} and the co-crystallized query poses were taken from the same site. The screening library contains 2{,}376 actives and 134{,}437 decoys summed over the 43 targets (2{,}298 and 70{,}526 distinct compounds, because some compounds appear in more than one target's library). Each compound is represented by up to 50 conformers embedded with RDKit ETKDG and pruned at 0.5\,\AA{} RMSD during embedding, then relaxed with MMFF94 where parameterization succeeds. Aromaticity is restored with \texttt{Chem.SanitizeMol} before featurization. Pruning leaves an average of 23.1 conformers per compound and about 3.2 million conformers in total.
For \sss{}, partial charges are computed with GFN2-xTB, molecular surfaces are represented by 200 sampled points, and pharmacophores are computed for each conformer. Surface sampling uses a fixed seed. The surface coordinates and corresponding electrostatic potentials are stored and reused for scoring. Each library compound is assigned its maximum similarity to the crystal-ligand query across its conformer ensemble. All \sss{} modes and open-source baselines utilize the same set of conformers.

The ESP-Sim\cite{bolcato2022espsim} baseline aligns each conformer with RDKit's \texttt{GetCrippenO3A}, then evaluates \texttt{GetEspSim} with analytic Gaussian integration and the default Carbo metric. Explicit hydrogens are added, and MMFF94 partial charges are used. The comparison uses the ESP score for retrieval, compared with \texttt{vol\_and\_surf\_esp}. ESP-Sim measures analytic electrostatic field similarity whereas \texttt{vol\_and\_surf\_esp} combines volumetric shape and sampled-surface ESP agreement.

From these rankings, we compute tie-aware ROC-AUC, 1\% enrichment factor (EF1\%), and the hit rate. The metrics are tie-aware because resolving ties at the top 1\% cutoff by input order can inflate EF1\% because actives are listed before decoys in the input. ROC-AUC measures active--decoy ordering across the full ranked list and does not specifically emphasize early retrieval, which EF1\% and hit rate measure.\cite{jain2008recommendations} EF1\% measures the fraction of actives in the top 1\% of the ranked library relative to random expectation, such that a random ranking has an expected EF1\% of $1.0$. Its maximum depends on library composition:
\begin{equation}
  \mathrm{EF1\%}_{\max} \;=\; \frac{\min(n_{\mathrm{top}},\,N_{\mathrm{act}})/n_{\mathrm{top}}}
                                   {N_{\mathrm{act}}/N},
  \qquad n_{\mathrm{top}} = \max\!\left(1,\ \mathrm{round}(0.01\,N)\right).
\end{equation}
Across these 43 targets, this maximum ranges from 29.4 to 79.8 because the libraries contain 23--95 actives against 1{,}135--6{,}116 property-matched decoys. Consequently, averaging raw EF1\% across targets conflates retrieval quality with library composition. To normalize the maximum attainable value, we also compute hit rate, $\mathrm{EF1\%}/\mathrm{EF1\%}_{\max}$, which equals $1.0$ for perfect retrieval regardless of library composition and has an expected value of $1/\mathrm{EF1\%}_{\max}$ under random ranking (approximately 0.013--0.034 here). This relative-enrichment measure has also been reported alongside EF in comparative assessments of 3D similarity methods.\cite{jiangComprehensiveComparativeAssessment2021} Results for EF1\% and hit rate are shown in Table~\ref{tab:si-modes}.

\subsubsection{Enrichment of the additional modes}
\label{sec:si-accuracy}

Table~\ref{tab:si-modes} reports enrichment for the ten additional modes and compares each Tversky variant with its corresponding Tanimoto mode.

\begin{table}[H]
\centering
\caption{Enrichment across all 43 DUDE-Z targets for all eighteen \sss{} modes and the external tool settings. Values are means $\pm$ SEM across targets. Hit rate is normalized EF1\%. For each Tversky mode, the last two columns give the ROC-AUC of its Tanimoto counterpart and the Tversky-minus-Tanimoto difference. Superscripts compare each metric with the mode listed in the Reference column using an unadjusted two-sided paired $t$-test over targets ($^{**}p<0.005$, $^{*}p<0.05$, n.s. not significant). A dash indicates no paired comparison.}
\label{tab:si-modes}
\small
\fitsitable{\begin{tabular}{llrrrrr}
\toprule
Method & Reference & ROC-AUC\makebox[1.6em][l]{} & EF1\%\makebox[1.6em][l]{} & Hit rate\makebox[1.6em][l]{} & Tanimoto ROC-AUC & $\Delta$ \\
\midrule
\multicolumn{7}{l}{\textbf{\sss{}}} \\
vol & -- & $0.657\pm 0.025$\makebox[1.6em][l]{} & $9.92\pm 1.32$\makebox[1.6em][l]{} & $0.175\pm 0.024$\makebox[1.6em][l]{} & -- & -- \\
vol\_esp & -- & $0.697\pm 0.026$\makebox[1.6em][l]{} & $13.18\pm 1.73$\makebox[1.6em][l]{} & $0.236\pm 0.031$\makebox[1.6em][l]{} & -- & -- \\
surf & -- & $0.657\pm 0.026$\makebox[1.6em][l]{} & $8.43\pm 1.06$\makebox[1.6em][l]{} & $0.147\pm 0.019$\makebox[1.6em][l]{} & -- & -- \\
surf\_esp & -- & $0.657\pm 0.026$\makebox[1.6em][l]{} & $8.12\pm 1.06$\makebox[1.6em][l]{} & $0.141\pm 0.019$\makebox[1.6em][l]{} & -- & -- \\
vol\_and\_surf\_esp & -- & $0.637\pm 0.025$\makebox[1.6em][l]{} & $11.50\pm 1.70$\makebox[1.6em][l]{} & $0.202\pm 0.030$\makebox[1.6em][l]{} & -- & -- \\
pharm & -- & $0.701\pm 0.023$\makebox[1.6em][l]{} & $15.11\pm 1.97$\makebox[1.6em][l]{} & $0.265\pm 0.034$\makebox[1.6em][l]{} & -- & -- \\
vol\_color & -- & $0.721\pm 0.022$\makebox[1.6em][l]{} & $16.48\pm 1.93$\makebox[1.6em][l]{} & $0.290\pm 0.034$\makebox[1.6em][l]{} & -- & -- \\
vol\_lipo & -- & $0.709\pm 0.025$\makebox[1.6em][l]{} & $13.69\pm 1.69$\makebox[1.6em][l]{} & $0.245\pm 0.031$\makebox[1.6em][l]{} & -- & -- \\
vol\_pharm & -- & $0.718\pm 0.022$\makebox[1.6em][l]{} & $15.95\pm 1.85$\makebox[1.6em][l]{} & $0.281\pm 0.033$\makebox[1.6em][l]{} & -- & -- \\
vol\_atomtype & -- & $0.702\pm 0.025$\makebox[1.6em][l]{} & $14.15\pm 1.73$\makebox[1.6em][l]{} & $0.254\pm 0.032$\makebox[1.6em][l]{} & -- & -- \\
vol\_mr & -- & $0.710\pm 0.024$\makebox[1.6em][l]{} & $15.25\pm 1.84$\makebox[1.6em][l]{} & $0.275\pm 0.036$\makebox[1.6em][l]{} & -- & -- \\
surf\_tversky & surf & $0.506\pm 0.023$\makebox[1.6em][l]{$^{**}$} & $3.73\pm 0.62$\makebox[1.6em][l]{$^{**}$} & $0.064\pm 0.010$\makebox[1.6em][l]{$^{**}$} & 0.657 & $-0.151$ \\
surf\_esp\_tversky & surf\_esp & $0.509\pm 0.023$\makebox[1.6em][l]{$^{**}$} & $4.00\pm 0.70$\makebox[1.6em][l]{$^{**}$} & $0.068\pm 0.011$\makebox[1.6em][l]{$^{**}$} & 0.657 & $-0.148$ \\
vol\_and\_surf\_esp\_tversky & vol\_and\_surf\_esp & $0.712\pm 0.019$\makebox[1.6em][l]{$^{**}$} & $13.40\pm 1.81$\makebox[1.6em][l]{$^{*}$} & $0.237\pm 0.033$\makebox[1.6em][l]{$^{*}$} & 0.637 & $+0.075$ \\
vol\_color\_tversky & vol\_color & $0.801\pm 0.016$\makebox[1.6em][l]{$^{**}$} & $18.63\pm 1.99$\makebox[1.6em][l]{$^{\mathrm{n.s.}}$} & $0.324\pm 0.036$\makebox[1.6em][l]{$^{\mathrm{n.s.}}$} & 0.721 & $+0.080$ \\
vol\_lipo\_tversky & vol\_lipo & $0.787\pm 0.014$\makebox[1.6em][l]{$^{**}$} & $15.71\pm 1.83$\makebox[1.6em][l]{$^{\mathrm{n.s.}}$} & $0.278\pm 0.033$\makebox[1.6em][l]{$^{\mathrm{n.s.}}$} & 0.709 & $+0.078$ \\
pharm\_tversky & pharm & $0.755\pm 0.020$\makebox[1.6em][l]{$^{**}$} & $16.36\pm 2.11$\makebox[1.6em][l]{$^{\mathrm{n.s.}}$} & $0.284\pm 0.037$\makebox[1.6em][l]{$^{\mathrm{n.s.}}$} & 0.701 & $+0.054$ \\
vol\_fukui & -- & $0.659\pm 0.025$\makebox[1.6em][l]{} & $9.82\pm 1.32$\makebox[1.6em][l]{} & $0.173\pm 0.024$\makebox[1.6em][l]{} & -- & -- \\
\midrule
\multicolumn{7}{l}{\textbf{External implementations}} \\
espsim\_esp & vol\_and\_surf\_esp & $0.652\pm 0.021$\makebox[1.6em][l]{$^{\mathrm{n.s.}}$} & $9.88\pm 2.45$\makebox[1.6em][l]{$^{\mathrm{n.s.}}$} & $0.171\pm 0.041$\makebox[1.6em][l]{$^{\mathrm{n.s.}}$} & -- & -- \\
pharao & pharm & $0.667\pm 0.024$\makebox[1.6em][l]{$^{*}$} & $11.33\pm 1.62$\makebox[1.6em][l]{$^{**}$} & $0.198\pm 0.029$\makebox[1.6em][l]{$^{**}$} & -- & -- \\
phesa\_pharm & pharm & $0.715\pm 0.022$\makebox[1.6em][l]{$^{\mathrm{n.s.}}$} & $14.50\pm 2.09$\makebox[1.6em][l]{$^{\mathrm{n.s.}}$} & $0.251\pm 0.033$\makebox[1.6em][l]{$^{\mathrm{n.s.}}$} & -- & -- \\
phesa\_shape & vol & $0.668\pm 0.025$\makebox[1.6em][l]{$^{*}$} & $10.63\pm 1.40$\makebox[1.6em][l]{$^{\mathrm{n.s.}}$} & $0.189\pm 0.025$\makebox[1.6em][l]{$^{*}$} & -- & -- \\
rdkit\_color & vol\_color & $0.699\pm 0.024$\makebox[1.6em][l]{$^{*}$} & $15.93\pm 1.89$\makebox[1.6em][l]{$^{\mathrm{n.s.}}$} & $0.284\pm 0.036$\makebox[1.6em][l]{$^{\mathrm{n.s.}}$} & -- & -- \\
rdkit\_shape & vol & $0.660\pm 0.025$\makebox[1.6em][l]{$^{\mathrm{n.s.}}$} & $10.11\pm 1.47$\makebox[1.6em][l]{$^{\mathrm{n.s.}}$} & $0.178\pm 0.026$\makebox[1.6em][l]{$^{\mathrm{n.s.}}$} & -- & -- \\
roshambo2\_combo & vol\_color & $0.705\pm 0.024$\makebox[1.6em][l]{$^{\mathrm{n.s.}}$} & $14.65\pm 1.83$\makebox[1.6em][l]{$^{\mathrm{n.s.}}$} & $0.267\pm 0.036$\makebox[1.6em][l]{$^{\mathrm{n.s.}}$} & -- & -- \\
roshambo2\_shape & vol & $0.654\pm 0.025$\makebox[1.6em][l]{$^{\mathrm{n.s.}}$} & $9.82\pm 1.30$\makebox[1.6em][l]{$^{\mathrm{n.s.}}$} & $0.173\pm 0.023$\makebox[1.6em][l]{$^{\mathrm{n.s.}}$} & -- & -- \\
shaep\_esp & vol\_and\_surf\_esp & $0.656\pm 0.024$\makebox[1.6em][l]{$^{\mathrm{n.s.}}$} & $10.73\pm 1.59$\makebox[1.6em][l]{$^{\mathrm{n.s.}}$} & $0.193\pm 0.030$\makebox[1.6em][l]{$^{\mathrm{n.s.}}$} & -- & -- \\
shaep\_shape & vol & $0.674\pm 0.025$\makebox[1.6em][l]{$^{*}$} & $10.56\pm 1.57$\makebox[1.6em][l]{$^{\mathrm{n.s.}}$} & $0.186\pm 0.028$\makebox[1.6em][l]{$^{\mathrm{n.s.}}$} & -- & -- \\
\bottomrule
\end{tabular}
}
\end{table}

The effect of Tversky normalization on retrieval depends on the molecular representation. Compared with their Tanimoto counterparts, \texttt{vol\_color\_tversky}, \texttt{vol\_and\_surf\_esp\_tversky}, \texttt{vol\_lipo\_tversky}, and \texttt{pharm\_tversky} improve mean ROC-AUC by 0.080, 0.075, 0.078, and 0.054, respectively. The asymmetric Tversky index can favor molecules that match the reference scaffold while allowing additional structural features. \texttt{vol\_color\_tversky} achieves the highest mean ROC-AUC of all methods evaluated at 0.801.

Surface-based representations perform worse with Tversky normalization. \texttt{surf\_tversky} and \texttt{surf\_esp\_tversky} achieve mean ROC-AUC values of 0.506 and 0.509, respectively, close to the random baseline of 0.5 and below 0.657 for both Tanimoto counterparts. Tversky normalization weights the self-overlaps $O_{AA}$ and $O_{BB}$ asymmetrically. These terms depend on molecular size and point density, which varies with surface area because each surface is represented by 200 points. This variation may contribute to the poorer retrieval, although its effect has not been isolated. Tanimoto normalization weights the self-overlaps equally but remains sensitive to molecular size and point density. The two surface-based Tversky variants also have the lowest mean EF1\% values among the evaluated methods at 3.7 and 4.0, respectively, compared with 1.0 for random ranking.

Notably, \texttt{vol\_fukui} is based on a per-atom reactivity field which was not designed as a binding-affinity predictor and serves as a demonstration that a new quantum-chemical descriptor can be added for fast alignment and scoring.

\subsubsection{Early-recognition enrichment across all methods}

Table~\ref{tab:si-modes} reports EF1\% and hit rate for all \sss{} and open-source methods, with per-target EF1\% values in Tables~\ref{tab:si-ef1-main} and \ref{tab:si-ef1-si}. Mean hit rate and mean EF1\% are strongly correlated across the twenty-eight methods (Spearman $\rho = 0.996$). Mean EF1\% and mean ROC-AUC are less strongly correlated ($\rho = 0.883$). Mean EF1\% across these methods varies substantially by target, from 1.6 for FGFR1 and 1.7 for LCK to 30.6 for FABP4 and 33.3 for CXCR4. Of the 1{,}204 method--target combinations, 146 fall below the random-ranking baseline of 1. No method achieves enrichment above this baseline for every target, indicating that early enrichment depends strongly on the target.

\begin{figure}[H]
  \centering
  \includegraphics[width=\linewidth]{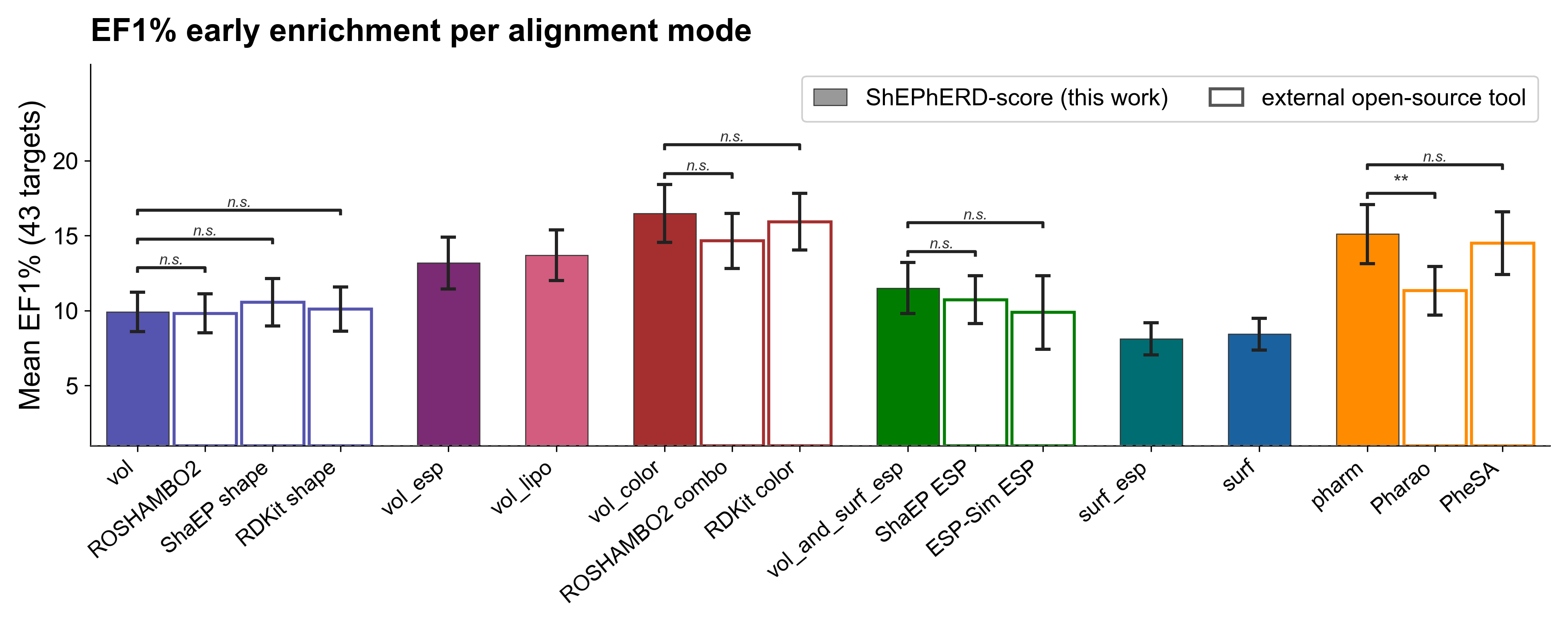}
  \caption{\textbf{Early enrichment on DUDE-Z.} Mean EF1\% across the 43 DUDE-Z targets for each \sss{} mode (solid) and the open-source implementation it is matched against (outlined, in the matched mode's color), plotted as in Figure~\ref{fig:enrichment}B. Bars start at the random-ranking baseline of EF1\% $=1$. Error bars are the standard error of the mean across targets. Brackets are unadjusted two-sided paired $t$-tests across targets ($^{**}p<0.005$, n.s.\ $=$ not significant).}
  \label{fig:si-ef1}
\end{figure}

Figure~\ref{fig:si-ef1} summarizes the nine matched comparisons of Figure~\ref{fig:enrichment}B using EF1\% in place of ROC-AUC. \texttt{pharm} reached a mean EF1\% of 15.11 compared with 11.33 for Pharao ($p=0.00054$, unadjusted two-sided paired $t$-test across 43 targets), and the remaining eight pairs were statistically indistinguishable. \sss{} therefore matched or exceeded the early-recognition performance of every comparator it was compared to.

\begin{table}[H]
\centering
\caption{Per-target EF1\% for all eighteen \sss{} modes. A random ranking has expected EF1\% of one.}
\label{tab:si-ef1-main}
\small
\fitsitable{\begin{tabular}{lp{2.6em}p{2.6em}p{2.6em}p{2.6em}p{2.6em}p{2.6em}p{2.6em}p{2.6em}p{2.6em}p{2.6em}p{2.6em}p{2.6em}p{2.6em}p{2.6em}p{2.6em}p{2.6em}p{2.6em}p{2.6em}}
\toprule
Target & \makebox[2.6em][l]{\rotatebox[origin=l]{65}{vol}} & \makebox[2.6em][l]{\rotatebox[origin=l]{65}{vol\_esp}} & \makebox[2.6em][l]{\rotatebox[origin=l]{65}{surf}} & \makebox[2.6em][l]{\rotatebox[origin=l]{65}{surf\_esp}} & \makebox[2.6em][l]{\rotatebox[origin=l]{65}{vol\_and\_surf\_esp}} & \makebox[2.6em][l]{\rotatebox[origin=l]{65}{pharm}} & \makebox[2.6em][l]{\rotatebox[origin=l]{65}{vol\_color}} & \makebox[2.6em][l]{\rotatebox[origin=l]{65}{vol\_lipo}} & \makebox[2.6em][l]{\rotatebox[origin=l]{65}{vol\_pharm}} & \makebox[2.6em][l]{\rotatebox[origin=l]{65}{vol\_atomtype}} & \makebox[2.6em][l]{\rotatebox[origin=l]{65}{vol\_mr}} & \makebox[2.6em][l]{\rotatebox[origin=l]{65}{surf\_tversky}} & \makebox[2.6em][l]{\rotatebox[origin=l]{65}{surf\_esp\_tversky}} & \makebox[2.6em][l]{\rotatebox[origin=l]{65}{vol\_and\_surf\_esp\_tversky}} & \makebox[2.6em][l]{\rotatebox[origin=l]{65}{vol\_color\_tversky}} & \makebox[2.6em][l]{\rotatebox[origin=l]{65}{vol\_lipo\_tversky}} & \makebox[2.6em][l]{\rotatebox[origin=l]{65}{pharm\_tversky}} & \makebox[2.6em][l]{\rotatebox[origin=l]{65}{vol\_fukui}} \\
\midrule
AA2AR & 3.4 & 11.5 & 2.3 & 2.3 & 14.9 & 21.8 & 14.9 & 10.3 & 17.2 & 11.5 & 16.0 & 1.1 & 2.3 & 16.0 & 19.5 & 9.2 & 26.4 & 4.6 \\
ABL1 & 6.1 & 2.0 & 4.1 & 6.1 & 2.0 & 10.2 & 12.2 & 8.1 & 12.2 & 6.1 & 6.1 & 0.0 & 0.0 & 6.1 & 20.4 & 8.1 & 14.3 & 6.1 \\
ACES & 3.6 & 4.8 & 2.4 & 3.6 & 7.2 & 6.0 & 7.2 & 4.8 & 7.2 & 6.0 & 9.6 & 3.6 & 3.6 & 7.2 & 8.4 & 4.8 & 6.0 & 3.6 \\
ADA & 12.8 & 19.1 & 8.0 & 8.0 & 36.7 & 14.3 & 25.5 & 27.1 & 22.3 & 25.5 & 23.9 & 3.2 & 4.8 & 31.9 & 20.7 & 23.9 & 15.9 & 14.3 \\
ADRB2 & 10.9 & 16.4 & 10.9 & 5.5 & 9.1 & 10.9 & 12.7 & 12.7 & 12.7 & 14.6 & 10.9 & 3.6 & 0.0 & 3.6 & 10.9 & 12.7 & 7.3 & 10.9 \\
AMPC & 18.7 & 27.0 & 16.6 & 16.6 & 12.5 & 16.6 & 25.0 & 31.2 & 25.0 & 29.1 & 31.2 & 8.3 & 4.2 & 10.4 & 16.6 & 25.0 & 16.6 & 18.7 \\
ANDR & 17.2 & 23.6 & 12.9 & 13.9 & 30.0 & 30.0 & 25.7 & 22.5 & 25.7 & 24.7 & 25.7 & 9.7 & 9.7 & 32.2 & 27.9 & 27.9 & 24.7 & 17.2 \\
CSF1R & 19.4 & 21.2 & 15.9 & 15.9 & 19.4 & 19.4 & 21.2 & 24.7 & 21.2 & 21.2 & 17.6 & 5.3 & 7.1 & 28.2 & 21.2 & 19.4 & 21.2 & 17.6 \\
CXCR4 & 33.8 & 43.9 & 27.0 & 27.0 & 33.8 & 33.8 & 37.1 & 37.1 & 37.1 & 40.5 & 33.8 & 16.9 & 20.3 & 30.4 & 43.9 & 47.3 & 33.8 & 33.8 \\
DEF & 21.1 & 25.0 & 21.1 & 21.1 & 15.4 & 34.6 & 44.2 & 26.9 & 42.3 & 28.8 & 30.7 & 13.4 & 13.4 & 15.4 & 36.5 & 28.8 & 32.7 & 23.1 \\
DRD4 & 4.1 & 12.3 & 12.3 & 12.3 & 12.3 & 16.4 & 12.3 & 12.3 & 8.2 & 8.2 & 16.4 & 0.0 & 4.1 & 12.3 & 12.3 & 20.5 & 8.2 & 4.1 \\
EGFR & 8.9 & 10.3 & 3.0 & 3.0 & 7.4 & 17.7 & 16.2 & 11.8 & 16.2 & 11.8 & 11.8 & 4.4 & 4.4 & 11.8 & 26.6 & 5.9 & 23.6 & 8.9 \\
FA10 & 8.8 & 4.4 & 4.4 & 4.4 & 4.4 & 4.4 & 8.8 & 0.0 & 8.8 & 8.8 & 0.0 & 0.0 & 4.4 & 4.4 & 4.4 & 4.4 & 0.0 & 8.8 \\
FA7 & 6.2 & 3.1 & 6.2 & 3.1 & 0.0 & 6.2 & 15.5 & 6.2 & 12.4 & 3.1 & 6.2 & 0.0 & 0.0 & 3.1 & 24.8 & 15.5 & 12.4 & 6.2 \\
FABP4 & 36.2 & 36.2 & 18.1 & 18.1 & 36.2 & 36.2 & 36.2 & 36.2 & 36.2 & 36.2 & 36.2 & 6.8 & 6.8 & 36.2 & 36.2 & 36.2 & 36.2 & 36.2 \\
FGFR1 & 0.0 & 0.0 & 0.0 & 0.0 & 2.2 & 2.2 & 2.2 & 0.0 & 2.2 & 0.0 & 0.0 & 0.0 & 0.0 & 4.4 & 13.3 & 2.2 & 15.5 & 0.0 \\
FKB1A & 2.2 & 6.5 & 21.8 & 24.0 & 15.3 & 2.2 & 6.5 & 10.9 & 8.7 & 4.4 & 4.4 & 4.4 & 4.4 & 10.9 & 2.2 & 17.4 & 0.0 & 2.2 \\
GLCM & 0.0 & 10.7 & 0.0 & 0.0 & 2.7 & 10.7 & 10.7 & 2.7 & 13.4 & 10.7 & 8.1 & 5.4 & 5.4 & 0.0 & 0.0 & 0.0 & 8.1 & 0.0 \\
HDAC8 & 8.0 & 15.9 & 4.0 & 4.0 & 13.3 & 23.9 & 35.8 & 19.9 & 23.9 & 15.9 & 19.9 & 1.3 & 1.3 & 13.3 & 35.8 & 14.6 & 25.2 & 9.3 \\
HIVPR & 6.6 & 11.0 & 15.4 & 17.6 & 11.0 & 8.8 & 8.8 & 11.0 & 8.8 & 11.0 & 8.8 & 11.0 & 15.4 & 13.2 & 6.6 & 6.6 & 0.0 & 6.6 \\
HMDH & 26.8 & 24.1 & 5.4 & 5.4 & 13.4 & 18.8 & 29.5 & 26.8 & 26.8 & 29.5 & 26.8 & 8.0 & 8.0 & 18.8 & 29.5 & 18.8 & 21.5 & 26.8 \\
HS90A & 4.1 & 2.0 & 4.1 & 2.0 & 2.0 & 14.2 & 8.1 & 4.1 & 8.1 & 2.0 & 8.1 & 0.0 & 0.0 & 6.1 & 34.5 & 4.1 & 38.5 & 4.1 \\
ITAL & 12.6 & 15.1 & 7.5 & 7.5 & 20.1 & 20.1 & 17.6 & 15.1 & 17.6 & 15.1 & 15.1 & 0.0 & 0.0 & 15.1 & 12.6 & 12.6 & 15.1 & 12.6 \\
KIT & 4.4 & 4.4 & 8.8 & 4.4 & 0.0 & 2.2 & 4.4 & 6.6 & 4.4 & 4.4 & 8.8 & 2.2 & 0.0 & 6.6 & 19.9 & 15.5 & 4.4 & 4.4 \\
KITH & 14.1 & 14.1 & 10.6 & 10.6 & 14.1 & 33.6 & 14.1 & 14.1 & 14.1 & 14.1 & 14.1 & 3.5 & 3.5 & 40.6 & 51.2 & 21.2 & 53.0 & 14.1 \\
LCK & 0.0 & 0.0 & 0.0 & 1.5 & 1.5 & 3.0 & 1.5 & 1.5 & 1.5 & 1.5 & 1.5 & 0.0 & 0.0 & 4.6 & 6.1 & 1.5 & 4.6 & 0.0 \\
MAPK2 & 19.9 & 37.1 & 11.9 & 7.9 & 22.5 & 23.8 & 34.4 & 34.4 & 33.1 & 33.1 & 39.7 & 9.3 & 7.9 & 29.1 & 34.4 & 39.7 & 21.2 & 19.9 \\
MK01 & 4.6 & 11.6 & 2.3 & 4.6 & 32.5 & 34.8 & 27.8 & 20.9 & 27.8 & 16.2 & 37.1 & 2.3 & 2.3 & 46.4 & 46.4 & 48.7 & 46.4 & 4.6 \\
MT1 & 14.1 & 17.7 & 3.5 & 3.5 & 35.4 & 60.1 & 49.5 & 21.2 & 49.5 & 17.7 & 28.3 & 7.1 & 10.6 & 28.3 & 31.8 & 17.7 & 49.5 & 17.7 \\
NRAM & 14.3 & 28.7 & 17.6 & 14.3 & 1.1 & 1.1 & 5.5 & 5.5 & 6.6 & 32.0 & 11.0 & 2.2 & 1.1 & 2.2 & 3.3 & 6.6 & 1.1 & 7.7 \\
PARP1 & 13.7 & 8.4 & 10.5 & 8.4 & 4.2 & 13.7 & 16.9 & 9.5 & 13.7 & 11.6 & 13.7 & 5.3 & 2.1 & 7.4 & 15.8 & 14.8 & 10.5 & 13.7 \\
PLK1 & 4.7 & 2.4 & 14.1 & 11.8 & 0.0 & 7.1 & 7.1 & 0.0 & 7.1 & 4.7 & 2.4 & 2.4 & 2.4 & 14.1 & 14.1 & 11.8 & 14.1 & 4.7 \\
PPARA & 12.7 & 19.0 & 19.0 & 19.0 & 16.9 & 23.2 & 29.6 & 23.2 & 25.3 & 23.2 & 25.3 & 4.2 & 4.2 & 8.4 & 21.1 & 19.0 & 19.0 & 12.7 \\
PTN1 & 2.4 & 2.4 & 1.2 & 0.0 & 1.2 & 1.2 & 3.7 & 3.7 & 3.7 & 3.7 & 3.7 & 0.0 & 0.0 & 2.4 & 0.0 & 0.0 & 0.0 & 2.4 \\
PUR2 & 9.8 & 17.1 & 2.4 & 2.4 & 7.3 & 17.1 & 12.2 & 14.7 & 12.2 & 17.1 & 22.0 & 0.0 & 0.0 & 7.3 & 9.8 & 7.3 & 12.2 & 7.3 \\
RENI & 3.6 & 7.1 & 5.9 & 5.9 & 2.4 & 1.2 & 8.3 & 8.3 & 7.1 & 7.1 & 8.3 & 2.4 & 2.4 & 5.9 & 5.9 & 4.7 & 0.0 & 3.6 \\
ROCK1 & 0.0 & 2.4 & 0.0 & 0.0 & 0.0 & 2.4 & 2.4 & 2.4 & 2.4 & 2.4 & 0.0 & 0.0 & 0.0 & 2.4 & 9.5 & 14.3 & 2.4 & 0.0 \\
SRC & 4.6 & 4.6 & 6.2 & 4.6 & 6.2 & 10.8 & 9.3 & 6.2 & 12.3 & 4.6 & 10.8 & 3.1 & 3.1 & 9.3 & 9.3 & 12.3 & 17.0 & 4.6 \\
THRB & 4.5 & 0.0 & 4.5 & 6.8 & 6.8 & 0.0 & 0.0 & 0.0 & 0.0 & 0.0 & 0.0 & 0.0 & 0.0 & 4.5 & 4.5 & 0.0 & 6.8 & 4.5 \\
TRY1 & 8.8 & 7.0 & 12.3 & 12.3 & 5.3 & 0.0 & 8.8 & 12.3 & 8.8 & 8.8 & 10.5 & 1.8 & 3.5 & 7.0 & 14.1 & 14.1 & 3.5 & 7.0 \\
TRYB1 & 0.0 & 0.0 & 0.0 & 0.0 & 0.0 & 5.4 & 0.0 & 0.0 & 0.0 & 0.0 & 0.0 & 0.0 & 0.0 & 0.0 & 8.0 & 18.8 & 18.8 & 0.0 \\
UROK & 4.1 & 2.7 & 1.4 & 2.7 & 0.0 & 8.2 & 8.2 & 8.2 & 6.8 & 6.8 & 8.2 & 0.0 & 0.0 & 0.0 & 13.6 & 9.5 & 6.8 & 4.1 \\
XIAP & 14.7 & 33.4 & 6.7 & 6.7 & 16.0 & 21.4 & 30.7 & 33.4 & 34.8 & 34.8 & 42.8 & 8.0 & 9.4 & 18.7 & 17.4 & 32.1 & 9.4 & 13.4 \\
\bottomrule
\end{tabular}
}
\end{table}

\begin{table}[H]
\centering
\caption{Per-target EF1\% for all external-tool settings. A random ranking has expected EF1\% of one.}
\label{tab:si-ef1-si}
\small
\fitsitable{\begin{tabular}{lp{2.6em}p{2.6em}p{2.6em}p{2.6em}p{2.6em}p{2.6em}p{2.6em}p{2.6em}p{2.6em}p{2.6em}}
\toprule
Target & \makebox[2.6em][l]{\rotatebox[origin=l]{65}{espsim\_esp}} & \makebox[2.6em][l]{\rotatebox[origin=l]{65}{pharao}} & \makebox[2.6em][l]{\rotatebox[origin=l]{65}{phesa\_pharm}} & \makebox[2.6em][l]{\rotatebox[origin=l]{65}{phesa\_shape}} & \makebox[2.6em][l]{\rotatebox[origin=l]{65}{rdkit\_color}} & \makebox[2.6em][l]{\rotatebox[origin=l]{65}{rdkit\_shape}} & \makebox[2.6em][l]{\rotatebox[origin=l]{65}{roshambo2\_combo}} & \makebox[2.6em][l]{\rotatebox[origin=l]{65}{roshambo2\_shape}} & \makebox[2.6em][l]{\rotatebox[origin=l]{65}{shaep\_esp}} & \makebox[2.6em][l]{\rotatebox[origin=l]{65}{shaep\_shape}} \\
\midrule
AA2AR & 10.3 & 13.8 & 19.5 & 3.4 & 10.3 & 3.4 & 8.0 & 4.6 & 10.3 & 8.0 \\
ABL1 & 8.1 & 8.1 & 4.1 & 4.1 & 6.1 & 4.1 & 6.1 & 4.1 & 2.0 & 4.1 \\
ACES & 4.8 & 7.2 & 7.2 & 4.8 & 8.4 & 3.6 & 6.0 & 3.6 & 6.0 & 4.8 \\
ADA & 44.6 & 12.8 & 43.0 & 11.2 & 41.5 & 11.2 & 28.7 & 14.3 & 19.1 & 19.1 \\
ADRB2 & 7.3 & 16.4 & 10.9 & 12.7 & 20.0 & 12.7 & 14.6 & 10.9 & 9.1 & 9.1 \\
AMPC & 0.0 & 8.3 & 14.6 & 18.7 & 25.0 & 20.8 & 31.2 & 18.7 & 12.5 & 14.6 \\
ANDR & 20.4 & 17.2 & 29.0 & 17.2 & 29.0 & 17.2 & 24.7 & 17.2 & 25.7 & 24.7 \\
CSF1R & 17.6 & 8.8 & 8.8 & 17.6 & 17.6 & 19.4 & 22.9 & 19.4 & 17.6 & 19.4 \\
CXCR4 & 3.4 & 23.6 & 33.8 & 40.5 & 33.8 & 40.5 & 33.8 & 33.8 & 37.1 & 40.5 \\
DEF & 13.4 & 13.4 & 17.3 & 23.1 & 36.5 & 26.9 & 42.3 & 23.1 & 25.0 & 26.9 \\
DRD4 & 16.4 & 12.3 & 16.4 & 8.2 & 12.3 & 8.2 & 8.2 & 4.1 & 12.3 & 12.3 \\
EGFR & 8.9 & 10.3 & 5.9 & 8.9 & 17.7 & 8.9 & 10.3 & 10.3 & 4.4 & 7.4 \\
FA10 & 0.0 & 0.0 & 4.4 & 13.3 & 4.4 & 8.8 & 0.0 & 8.8 & 4.4 & 8.8 \\
FA7 & 0.0 & 6.2 & 3.1 & 6.2 & 9.3 & 3.1 & 6.2 & 6.2 & 3.1 & 3.1 \\
FABP4 & 0.0 & 31.7 & 11.3 & 36.2 & 36.2 & 36.2 & 38.5 & 36.2 & 36.2 & 36.2 \\
FGFR1 & 0.0 & 0.0 & 0.0 & 0.0 & 0.0 & 0.0 & 0.0 & 0.0 & 0.0 & 0.0 \\
FKB1A & 8.7 & 6.5 & 4.4 & 6.5 & 13.1 & 2.2 & 2.2 & 2.2 & 4.4 & 2.2 \\
GLCM & 2.7 & 10.7 & 18.8 & 0.0 & 2.7 & 0.0 & 8.1 & 0.0 & 0.0 & 2.7 \\
HDAC8 & 21.2 & 14.6 & 34.5 & 8.0 & 34.5 & 8.0 & 23.9 & 8.0 & 5.3 & 5.3 \\
HIVPR & 0.0 & 8.8 & 6.6 & 11.0 & 8.8 & 8.8 & 8.8 & 6.6 & 6.6 & 15.4 \\
HMDH & 2.7 & 2.7 & 13.4 & 26.8 & 24.1 & 29.5 & 13.4 & 24.1 & 29.5 & 32.2 \\
HS90A & 2.0 & 10.1 & 30.4 & 4.1 & 12.2 & 4.1 & 4.1 & 4.1 & 0.0 & 0.0 \\
ITAL & 5.0 & 20.1 & 5.0 & 15.1 & 17.6 & 15.1 & 17.6 & 12.6 & 17.6 & 17.6 \\
KIT & 0.0 & 2.2 & 2.2 & 4.4 & 4.4 & 2.2 & 8.8 & 4.4 & 4.4 & 6.6 \\
KITH & 51.2 & 37.1 & 17.7 & 12.4 & 17.7 & 12.4 & 14.1 & 14.1 & 14.1 & 14.1 \\
LCK & 0.0 & 7.6 & 0.0 & 0.0 & 4.6 & 0.0 & 1.5 & 0.0 & 3.0 & 0.0 \\
MAPK2 & 39.7 & 18.5 & 18.5 & 25.2 & 30.5 & 21.2 & 34.4 & 19.9 & 18.5 & 30.5 \\
MK01 & 46.4 & 41.7 & 32.5 & 7.0 & 44.1 & 4.6 & 39.4 & 4.6 & 34.8 & 9.3 \\
MT1 & 67.2 & 35.4 & 70.7 & 14.1 & 21.2 & 14.1 & 17.7 & 7.1 & 7.1 & 3.5 \\
NRAM & 0.0 & 0.0 & 13.2 & 9.9 & 8.8 & 13.2 & 16.5 & 13.2 & 19.9 & 12.1 \\
PARP1 & 5.3 & 13.7 & 9.5 & 8.4 & 13.7 & 9.5 & 9.5 & 12.7 & 6.3 & 7.4 \\
PLK1 & 0.0 & 0.0 & 2.4 & 9.4 & 2.4 & 9.4 & 7.1 & 4.7 & 0.0 & 4.7 \\
PPARA & 2.1 & 23.2 & 19.0 & 12.7 & 25.3 & 10.6 & 23.2 & 12.7 & 6.3 & 4.2 \\
PTN1 & 1.2 & 0.0 & 4.9 & 2.4 & 1.2 & 2.4 & 2.4 & 2.4 & 0.0 & 1.2 \\
PUR2 & 2.4 & 7.3 & 12.2 & 12.2 & 12.2 & 9.8 & 24.5 & 7.3 & 14.7 & 4.9 \\
RENI & 4.7 & 0.0 & 4.7 & 4.7 & 9.5 & 2.4 & 7.1 & 3.6 & 4.7 & 8.3 \\
ROCK1 & 0.0 & 0.0 & 2.4 & 0.0 & 2.4 & 0.0 & 2.4 & 0.0 & 0.0 & 0.0 \\
SRC & 1.5 & 3.1 & 6.2 & 4.6 & 12.3 & 3.1 & 3.1 & 4.6 & 6.2 & 4.6 \\
THRB & 0.0 & 2.3 & 6.8 & 2.3 & 0.0 & 4.5 & 0.0 & 9.1 & 2.3 & 2.3 \\
TRY1 & 0.0 & 1.8 & 8.8 & 7.0 & 5.3 & 5.3 & 10.5 & 8.8 & 5.3 & 7.0 \\
TRYB1 & 0.0 & 2.7 & 10.7 & 0.0 & 2.7 & 0.0 & 5.4 & 0.0 & 0.0 & 0.0 \\
UROK & 0.0 & 5.4 & 9.5 & 5.4 & 9.5 & 2.7 & 12.2 & 5.4 & 8.2 & 4.1 \\
XIAP & 5.3 & 21.4 & 29.4 & 17.4 & 36.1 & 14.7 & 30.7 & 14.7 & 17.4 & 14.7 \\
\bottomrule
\end{tabular}
}
\end{table}

\subsubsection{Charge-model comparison}

Table~\ref{tab:si-charge-comparison} compares enrichment using the stored GFN2-xTB charges with MMFF94 charges for the same query and library conformers. MMFF94 charges were obtained from the hydrogen-containing RDKit \texttt{Molecule} object \texttt{mol} using
\begin{verbatim}
props = AllChem.MMFFGetMoleculeProperties(Chem.Mol(mol), mmffVariant='MMFF94')
charges = np.array(
    [props.GetMMFFPartialCharge(i) for i in range(mol.GetNumAtoms())],
    dtype=np.float32
)
\end{verbatim}
The xTB single-point charge calculation uses the command
\begin{verbatim}
xtb input_mol.xyz --scc --parallel 1 --chrg 0
\end{verbatim}
with GFN2-xTB as the default method and no implicit solvent. The preparation wrapper retries failed SCC calculations with \texttt{--iterations 1000}. MMFF94 retains the molecular formal charge, whereas the stored xTB calculation specifies total charge zero. Surface ESP values are recomputed from the replacement partial charges at the frozen surface points; coordinates and all non-charge features are unchanged.

\begin{table}[H]
\centering
\caption{Charge-model comparison across 43 targets for the ESP-based modes. Values are means $\pm$ SEM; $\Delta$ rows are MMFF94 minus stored GFN2-xTB. Coordinates, surface points, non-charge features, and alignment settings are held fixed. The stored xTB charges follow the original ShEPhERD-score xTB interface, which defaults to total charge zero, whereas MMFF94 charges sum to the molecular formal charge. ROC-AUC did not change significantly for the three ESP-based modes in the main text, \texttt{vol\_esp}, \texttt{surf\_esp}, and \texttt{vol\_and\_surf\_esp}. The reported $p$-values are from unadjusted two-sided paired $t$-tests across targets.}
\label{tab:si-charge-comparison}
\small
\fitsitable{\begin{tabular}{llrrr}
\toprule
Mode & Quantity & \multicolumn{3}{c}{Enrichment metric} \\
\cmidrule(lr){3-5} & & ROC-AUC & EF1\% & Hit rate \\
\midrule
vol\_esp & xTB & $0.697\pm 0.026$ & $13.176\pm 1.731$ & $0.236\pm 0.031$ \\
 & MMFF94 & $0.715\pm 0.023$ & $15.390\pm 2.047$ & $0.280\pm 0.040$ \\
 & $\Delta$ & $+0.017\pm 0.015$ & $+2.215\pm 1.282$ & $+0.044\pm 0.024$ \\
 & Paired $p$ & 0.266 & 0.0914 & 0.0722 \\
\midrule
surf\_esp & xTB & $0.657\pm 0.026$ & $8.119\pm 1.063$ & $0.141\pm 0.019$ \\
 & MMFF94 & $0.655\pm 0.023$ & $8.335\pm 1.243$ & $0.146\pm 0.022$ \\
 & $\Delta$ & $-0.002\pm 0.006$ & $+0.216\pm 0.888$ & $+0.005\pm 0.014$ \\
 & Paired $p$ & 0.744 & 0.809 & 0.731 \\
\midrule
vol\_and\_surf\_esp & xTB & $0.637\pm 0.025$ & $11.504\pm 1.704$ & $0.202\pm 0.030$ \\
 & MMFF94 & $0.638\pm 0.020$ & $8.814\pm 1.579$ & $0.155\pm 0.028$ \\
 & $\Delta$ & $+0.001\pm 0.013$ & $-2.690\pm 1.377$ & $-0.048\pm 0.024$ \\
 & Paired $p$ & 0.927 & 0.0575 & 0.0487 \\
\midrule
vol\_esp\_tversky & xTB & $0.786\pm 0.014$ & $13.861\pm 1.699$ & $0.246\pm 0.030$ \\
 & MMFF94 & $0.789\pm 0.015$ & $18.505\pm 2.172$ & $0.333\pm 0.041$ \\
 & $\Delta$ & $+0.004\pm 0.011$ & $+4.644\pm 1.419$ & $+0.088\pm 0.026$ \\
 & Paired $p$ & 0.738 & 0.00213 & 0.00196 \\
\midrule
surf\_esp\_tversky & xTB & $0.509\pm 0.023$ & $4.001\pm 0.705$ & $0.068\pm 0.011$ \\
 & MMFF94 & $0.536\pm 0.020$ & $3.393\pm 0.649$ & $0.057\pm 0.010$ \\
 & $\Delta$ & $+0.027\pm 0.008$ & $-0.607\pm 0.584$ & $-0.010\pm 0.008$ \\
 & Paired $p$ & 0.00146 & 0.304 & 0.209 \\
\midrule
vol\_and\_surf\_esp\_tversky & xTB & $0.712\pm 0.019$ & $13.405\pm 1.813$ & $0.237\pm 0.033$ \\
 & MMFF94 & $0.702\pm 0.015$ & $11.090\pm 1.808$ & $0.193\pm 0.032$ \\
 & $\Delta$ & $-0.010\pm 0.012$ & $-2.315\pm 1.422$ & $-0.044\pm 0.024$ \\
 & Paired $p$ & 0.41 & 0.111 & 0.0741 \\
\bottomrule
\end{tabular}
}
\end{table}

\clearpage
\subsection{Alignment throughput and scaling}
\label{sec:si-throughput}

\subsubsection{Benchmark libraries and on-disk stores}
\label{sec:si-libraries}

To measure throughput, we sampled 210{,}000 compounds from the shuffled 13.6-million-compound Enamine REAL Sample subset (downloaded August 2026). We generated 50 ETKDG conformers per compound without RMS pruning and optimized each with MMFF94 for 200 iterations. Compounds yielding fewer than 50 conformers were excluded. Each conformer was treated as a separate library entry and prepared with MMFF94 partial charges, 200 surface points, and pharmacophore features, giving a pool of 10{,}497{,}450 conformers.

Throughput measurements used MMFF94 charges, whereas the primary \sss{} enrichment results used GFN2-xTB charges. Enrichment results with MMFF94 charges are reported in Table~\ref{tab:si-charge-comparison}, and GFN2-xTB preparation time is included in Figure~\ref{fig:speed}A.

Screening libraries were constructed by sampling $\lceil N/50\rceil$ compounds without replacement, including all their conformers, and truncating to $N$ conformers. Screens of $10^{8}$ conformers used an expanded pool of 104{,}978{,}250 conformers prepared from 2.1 million input compounds. Table~\ref{tab:benchmark-sizes} summarizes the workloads.

\begin{table}[H]
\centering
\caption{Benchmark sizes. $N$ is the number of library conformers for screening and conformer pairs for pairwise alignment. Enrichment uses 43 targets; preparation uses 100 compounds with 50 conformers each. Every timing measurement uses one untimed warm-up and one timed run.}
\label{tab:benchmark-sizes}
\small
\fitsitable{\begin{tabular}{lllllp{6cm}}
\toprule
Tool & Backend & Workload & Workers & GPUs & $N$ \\
\midrule
ESP-Sim & cpu & screen & 1 & 0 & 1,000 \\
\sss{} & numba & pairwise & 1 & 0 & 1,000 \\
\sss{} & numba & parallel & 1 & 0 & 10,000 \\
\sss{} & numba & parallel & 16 & 0 & 10,000 \\
\sss{} & numba & parallel & 32 & 0 & 10,000 \\
\sss{} & numba & parallel & 64 & 0 & 10,000 \\
\sss{} & numba & parallel & 8 & 0 & 10,000 \\
\sss{} & numba & screen & 1 & 0 & 10, 100, 1,000, 10,000, 100,000 \\
\sss{} & triton & pairwise & 1 & 1 & 100,000 \\
\sss{} & triton & screen & 1 & 1 & 10, 100, 1,000, 10,000, 100,000, 1,000,000, 10,000,000, 100,000,000 \\
\sss{} & triton & screen & 1 & 2 & 100,000,000 \\
\sss{} & triton & screen & 1 & 4 & 100,000,000 \\
Pharao & cpu & screen & 1 & 0 & 1,000 \\
PheSA & cpu & screen & 1 & 0 & 1,000 \\
RDKit & cpu & screen & 1 & 0 & 1,000 \\
ROSHAMBO2 & cuda & pairwise & 1 & 1 & 100,000 \\
ROSHAMBO2 & cuda & screen & 1 & 1 & 100,000, 10,000,000 \\
ShaEP & cpu & screen & 1 & 0 & 1,000 \\
\bottomrule
\end{tabular}
}
\end{table}

CPU screening was measured through $N=10^5$ for volumetric and pharmacophore modes and $N=10^3$ for modes requiring surfaces. All GPU pairwise measurements used the same 250 query conformers and 400 library conformers, with one conformer per compound and no compounds shared between the two sets. Each run evaluated all $250\times400=100{,}000$ pairs.

CPU screening comparisons used identical query and library conformers across tools, each restricted to one thread on an AMD EPYC 9474F. GPU screening and pairwise comparisons used a single NVIDIA L40S. Each measurement consisted of one untimed warm-up followed by one timed run.

Preparation times were measured on an AMD EPYC 9474F node using 100 compounds with a median of 26 heavy atoms (interquartile range 24--28), 6 rotatable bonds (5--8), and a molecular weight of 372\,Da (345--392). Shared preparation generated exactly 50 conformers per compound using ETKDGv3 with seed 42 and RMS pruning disabled, followed by MMFF94 relaxation and aromaticity restoration with \texttt{Chem.SanitizeMol}. All conversion measurements used the same saved set of 5{,}000 conformers. Shared preparation and conversion were timed separately, and their times were added to obtain total preparation time. Each was measured in one timed run. For \sss{} conversion, fresh molecule objects were used for both a full untimed warm-up and a timed pass over all 5{,}000 conformers.

Index construction was timed separately using prepared objects and included profile extraction and writing to a new directory. SDF export was also reported separately from tool-specific conversion because ShaEP, PheSA, and Pharao extract additional molecular features during alignment. ROSHAMBO2 preparation was measured separately for shape and shape-plus-color similarity.

\sss{} preparation used MMFF94 charges for \texttt{vol}, \texttt{pharm}, and \texttt{surf}, and GFN2-xTB charges for \texttt{vol\_esp} and \texttt{surf\_esp}. Pharmacophore extraction was disabled for \texttt{vol} and enabled for the other four preparation configurations. Surface representations used 200 points. Per-atom lipophilicity contributions were calculated when constructing molecular objects and added to the stored profiles during index construction. The \texttt{vol\_lipo} index was built from objects prepared using the \texttt{pharm} configuration, which took 317.4\,s including shared preparation. As a result, the time includes pharmacophore extraction although \texttt{vol\_lipo} does not require it.

\begin{table}[H]
\centering
\caption{Preparation times for 5{,}000 conformers from 100 compounds on one CPU core (mean 26.2 heavy atoms per compound). Conversion is the tool-specific preparation of the saved RDKit conformers, including the charges, surfaces, or pharmacophore features required by that configuration. Shared is the single conformer-generation run (ETKDGv3 embedding, MMFF94 relaxation, and aromaticity restoration; 308.5\,s, or 61.7\,ms per conformer) that every configuration reuses, so it is identical in every row; totals add it to each conversion time. Index construction is excluded and reported in Table~\ref{tab:si-index}.}
\label{tab:si-preparation-times}
\small
\fitsitable{\begin{tabular}{lrrr}
\toprule
Configuration & Shared (s) & Conversion (s) & Total (s) \\
\midrule
shared\_embedding & 308.547 & 0.000 & 308.547 \\
vol & 308.547 & 1.979 & 310.526 \\
pharm & 308.547 & 8.832 & 317.379 \\
surf & 308.547 & 624.271 & 932.818 \\
vol\_esp & 308.547 & 403.249 & 711.796 \\
surf\_esp & 308.547 & 1027.032 & 1335.579 \\
roshambo2\_shape & 308.547 & 3.754 & 312.301 \\
roshambo2\_color & 308.547 & 27.685 & 336.232 \\
sdf & 308.547 & 0.678 & 309.225 \\
\bottomrule
\end{tabular}
}
\par\end{table}
\begin{table}[H]
\centering
\caption{Index construction from the same 5{,}000 prepared conformers (100 compounds; mean 26.2 heavy atoms per compound). Construction extracts the mode-specific profiles and writes the screening store. Storage uses decimal megabytes ($1\,\mathrm{MB}=10^6$ bytes).}
\label{tab:si-index}
\small
\fitsitable{\begin{tabular}{lrrr}
\toprule
Mode & Conformers & Construction (s) & Storage (MB) \\
\midrule
vol & 5000 & 0.239 & 1.05 \\
pharm & 5000 & 0.311 & 2.05 \\
surf & 5000 & 0.299 & 7.05 \\
vol\_esp & 5000 & 0.308 & 2.14 \\
surf\_esp & 5000 & 0.329 & 9.05 \\
vol\_lipo & 5000 & 0.318 & 2.14 \\
\bottomrule
\end{tabular}
}
\end{table}

Table~\ref{tab:si-index} reports construction time and storage for the six measured indexes, each built from 5{,}000 conformers. These measurements are separate from the preparation totals in Table~\ref{tab:si-preparation-times}.
\subsubsection{Timing and alignment protocol}
\label{sec:si-protocol}

The throughput measurements in Figure~\ref{fig:speed} compare each package's default alignment workflow without isolating kernel execution or matching optimization effort. At its defaults, \sss{} performs up to 300 gradient evaluations per \texttt{vol} alignment (10 starting poses with 30 steps each), 640 for \texttt{vol\_color} (16 poses with 40 steps each), and 1{,}600 for \texttt{pharm} (32 poses with 50 steps each). ROSHAMBO2 uses four starting poses with 100 steps each for both scoring modes, giving up to 400 gradient evaluations per alignment.\cite{roshambo2_2025} For volumetric shape similarity, \sss{} allows 25\% fewer gradient evaluations and screens 9.1\% faster at $N=10^{7}$. For shape-plus-color similarity, it achieves 2.05$\times$ the throughput of ROSHAMBO2 despite allowing 60\% more gradient evaluations. RDKit's \texttt{rdShapeAlign} starts from an inertial-frame superposition and performs at most 10 pre-optimization and 30 post-optimization iterations. Pharao/Align-it superposes matched pharmacophore features by their centroids and then optimizes rotation only. Throughput therefore reflects both implementation efficiency and each package's balance between search effort and speed.

Early stopping can reduce the number of gradient evaluations below these limits. Where enabled in \sss{}, convergence is checked every five steps using a score-improvement tolerance of $10^{-5}$. CPU and GPU implementations require two consecutive checks below this tolerance for \texttt{vol}, \texttt{surf}, \texttt{vol\_color}, and \texttt{vol\_lipo}, and five for \texttt{vol\_esp}, \texttt{surf\_esp}, and \texttt{pharm}. Batched optimization stops only when all pairs meet the convergence criterion. Combined volume-and-surface electrostatic modes use all optimization steps in the CUDA-graph implementation. CPU and GPU scores may differ numerically despite identical stopping settings. Table~\ref{tab:si-alignment-settings} lists the starting poses, step limits, and scoring parameters for every mode.

\begin{table}[H]
\centering
\caption{Alignment settings used for the DUDE-Z benchmark, as recorded with the results. Every mode used the Adam optimizer with learning rate 0.1. Starts are initial orientations refined in parallel; evaluations are starts $\times$ steps, the maximum before early stopping. Patience is the number of consecutive five-step convergence checks below a score-improvement tolerance of $10^{-5}$ required to stop early. All modes except the pharmacophore modes use a Gaussian exponent $\alpha=0.81$; $\lambda$ is the width parameter of the electrostatic or property term as passed to the code, and for the surface ESP modes it is multiplied by a fixed scaling constant. Tversky modes use weights 0.95 and 0.05. Throughput measurements used the package defaults, which have the same starts and step limits.}
\label{tab:si-alignment-settings}
\fitsitable{\begin{tabular}{lrrrrl}
\toprule
Mode & Starts & Steps & Evaluations & Patience & Other settings \\
\midrule
\texttt{vol} & 10 & 30 & 300 & 2 & heavy atoms \\
\texttt{vol\_esp} & 16 & 50 & 800 & 5 & heavy atoms; $\lambda=0.1$ \\
\texttt{surf} & 8 & 40 & 320 & 2 & -- \\
\texttt{surf\_esp} & 8 & 40 & 320 & 5 & $\lambda=0.3$ \\
\texttt{vol\_and\_surf\_esp} & 8 & 60 & 480 & 5 & $\lambda=0.001$; probe radius 1.0\,\AA; ESP weight 0.5 \\
\texttt{pharm} & 32 & 50 & 1{,}600 & 5 & Tanimoto \\
\texttt{vol\_color} & 16 & 40 & 640 & 2 & color weight 0.5 \\
\texttt{vol\_lipo} & 16 & 50 & 800 & 2 & $\lambda=0.1$; lipophilicity weight 0.5 \\
\texttt{vol\_pharm} & 32 & 50 & 1{,}600 & 2 & pharmacophore weight 0.5 \\
\texttt{vol\_atomtype} & 16 & 40 & 640 & 2 & atom-type weight 0.5 \\
\texttt{vol\_mr} & 16 & 50 & 800 & 2 & $\lambda=0.1$; molar-refractivity weight 0.5 \\
\texttt{vol\_fukui} & 16 & 50 & 800 & 2 & $\lambda=0.1$; Fukui weight 0.5 \\
\texttt{surf\_tversky} & 8 & 40 & 320 & 2 & Tversky \\
\texttt{surf\_esp\_tversky} & 8 & 40 & 320 & 5 & $\lambda=0.3$; Tversky \\
\texttt{vol\_and\_surf\_esp\_tversky} & 8 & 60 & 480 & 5 & $\lambda=0.001$; probe radius 1.0\,\AA; ESP weight 0.5; Tversky \\
\texttt{vol\_color\_tversky} & 16 & 40 & 640 & 2 & color weight 0.5; Tversky \\
\texttt{vol\_lipo\_tversky} & 16 & 50 & 800 & 2 & $\lambda=0.1$; lipophilicity weight 0.5; Tversky \\
\texttt{pharm\_tversky} & 32 & 50 & 1{,}600 & 5 & Tversky \\
\bottomrule
\end{tabular}}
\end{table}

GPU measurements used a fixed set of Triton launch configurations for the recorded NVIDIA L40S hardware and Triton version. All GPU workers used the same configurations, whose hash was saved with each measurement. Runs requiring an unlisted configuration failed rather than initiating additional tuning. Host computation was limited to one thread per GPU worker, with I/O read-ahead threads enabled. The same query was used at every screening size. For \texttt{vol} screening of $10^8$ conformers on one, two, and four GPUs, the stored library was read before screening within each job. This initial read and the untimed warm-up were excluded from alignment time, so the reported rates measure screening after the initial library read rather than performance on first access from disk.

CPU measurements used Numba 0.59.1 and llvmlite 0.42.0 with Intel SVML vector math enabled, and whether SVML was active was recorded with every measurement. Numba 0.62 and later no longer include SVML support; without it the CPU kernels run more slowly, especially for the surface modes.

\subsubsection{Throughput of the additional modes}
\label{sec:si-additional-throughput}

GPU pairwise alignment throughput was measured on one NVIDIA L40S using the same 100{,}000 conformer pairs as the main-text comparison, drawn from the Enamine REAL pool described in Section~\ref{sec:si-libraries}. CPU pairwise measurements used the first 1{,}000 pairs from the same ordered $250\times400$ query--library combinations on one AMD EPYC 9474F core.

Table~\ref{tab:si-throughput} reports screening and pairwise throughput at $N=10^{5}$ on GPU and $N=10^{3}$ on one CPU core. All modes used the same selections of stored conformers. Each measurement consisted of one untimed warm-up followed by one timed run, excluding preparation and batch assembly.

\begin{table}[H]
\centering
\caption{Throughput in alignments per second. Columns give CPU screening and pairwise alignment at $N=10^3$, GPU screening at $N=10^5$, and GPU pairwise alignment at $N=10^5$. CPU measurements use one AMD EPYC 9474F core; GPU measurements use an NVIDIA L40S. Each rate comes from one timed run after one untimed warm-up. Dashes indicate unmeasured configurations. The main-text GPU screening comparison uses $N=10^7$; the common $N=10^5$ column here covers all modes. ROSHAMBO2 shape and color screening rates are measured at the same $N=10^5$ as the \sss{} modes.}
\label{tab:si-throughput}
\small
\fitsitable{\begin{tabular}{lrrrr}
\toprule
Tool / mode & \shortstack{CPU screen\\$N=10^3$} & \shortstack{CPU pairs\\$N=10^3$} & \shortstack{GPU screen\\$N=10^5$} & \shortstack{GPU pairs\\$N=10^5$} \\
\midrule
\sss{} / vol & 2,470.5 & 2,020.2 & 904,715.4 & 118,293.5 \\
\sss{} / vol\_esp & 841.6 & 797.4 & 363,972.1 & 82,823.3 \\
\sss{} / surf & 25.7 & 37.5 & 51,327.0 & 40,090.2 \\
\sss{} / surf\_esp & 28.2 & 34.7 & 32,510.2 & 25,862.9 \\
\sss{} / vol\_and\_surf\_esp & 19.0 & 19.4 & 18,049.4 & 13,400.6 \\
\sss{} / pharm & 574.1 & 515.2 & 165,546.4 & 63,657.9 \\
\sss{} / vol\_color & 813.0 & 682.0 & 332,361.3 & 60,727.6 \\
\sss{} / vol\_lipo & 468.4 & 409.8 & 215,117.6 & 58,492.9 \\
\sss{} / vol\_pharm & 280.0 & 234.0 & 98,648.2 & 39,251.2 \\
\sss{} / vol\_atomtype & 267.7 & 258.6 & 258,021.9 & 57,970.7 \\
\sss{} / vol\_mr & 323.2 & 294.3 & 214,050.2 & 57,154.2 \\
\sss{} / surf\_tversky & 20.6 & 37.5 & 38,052.7 & 31,135.9 \\
\sss{} / surf\_esp\_tversky & 19.6 & 34.7 & 32,471.7 & 25,846.3 \\
\sss{} / vol\_and\_surf\_esp\_tversky & 19.0 & 19.5 & 18,112.5 & 13,070.6 \\
\sss{} / vol\_color\_tversky & 789.8 & 678.6 & 331,031.5 & 61,816.8 \\
\sss{} / vol\_lipo\_tversky & 475.3 & 406.2 & 211,608.1 & 51,846.5 \\
\sss{} / pharm\_tversky & 572.6 & 510.0 & 163,817.9 & 62,641.5 \\
\sss{} / vol\_fukui & 461.7 & 409.7 & 213,585.2 & 58,100.6 \\
RDKit / vol & 726.2 & -- & -- & -- \\
RDKit / vol\_color & 666.5 & -- & -- & -- \\
PheSA / vol & 186.0 & -- & -- & -- \\
PheSA / pharm & 179.7 & -- & -- & -- \\
ShaEP / vol & 22.5 & -- & -- & -- \\
ShaEP / vol\_and\_surf\_esp & 22.5 & -- & -- & -- \\
Pharao / pharm & 1,533.4 & -- & -- & -- \\
ESP-Sim / vol\_and\_surf\_esp & 187.1 & -- & -- & -- \\
ROSHAMBO2 / vol & -- & -- & 887,879.9 & 7,415.4 \\
ROSHAMBO2 / vol\_color & -- & -- & 141,254.5 & 1,322.8 \\
\bottomrule
\end{tabular}
}
\end{table}
\clearpage

\subsubsection{Multi-process and multi-device screening}
\label{sec:si-multiproc}

\begin{figure}[htbp]
  \centering
  \includegraphics[width=\linewidth]{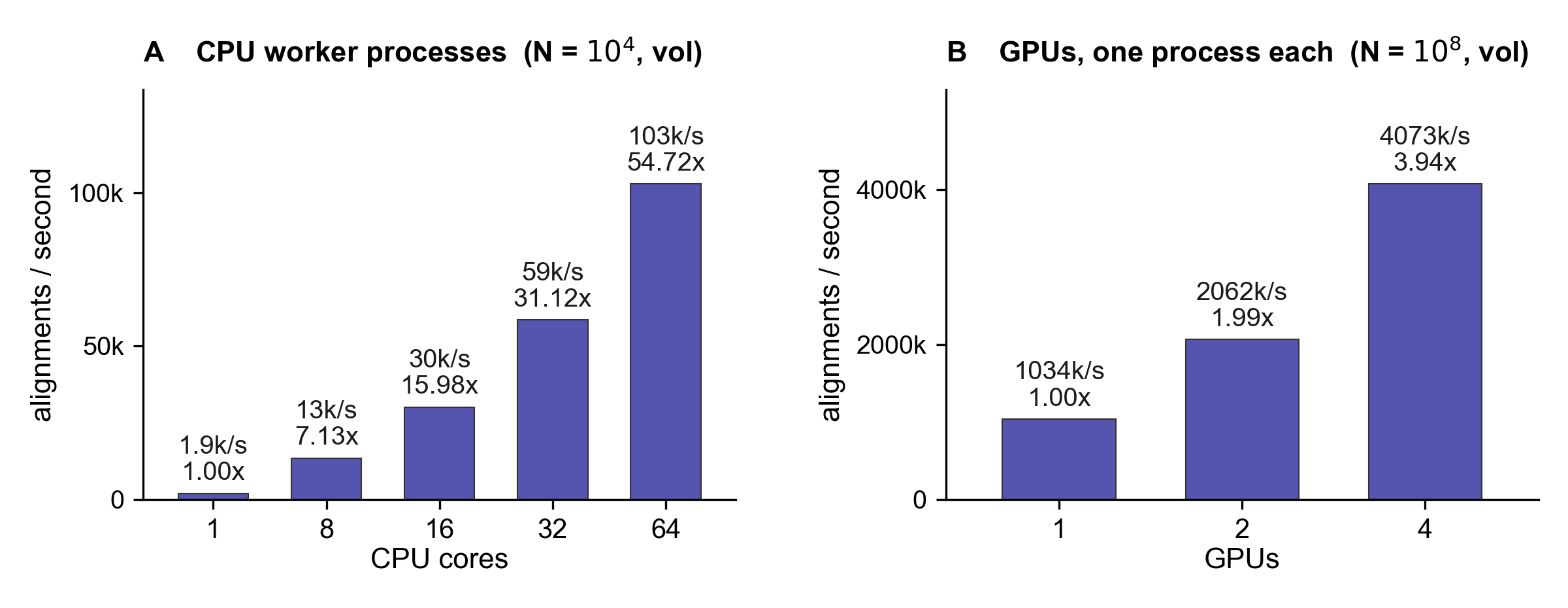}
  \caption{\texttt{vol} parallel screening throughput.
    \textbf{(A)} Parallelizing across CPUs for alignment of one query conformer against $N=10{,}000$ conformers. \textbf{(B)} Parallelizing across GPUs for alignment of one query conformer against $10^{8}$ conformers sharded across
    one, two, and four NVIDIA L40S GPUs. Labels show measured throughput and its ratio to the one-worker or one-GPU rate. Each rate uses one timed run after one warm-up; the ratios are observed wall-clock speedups, not prescribed linear scaling. Startup is excluded.}
  \label{fig:si-multiproc}
\end{figure}

The screening workflow distributes library entries across one worker process per CPU core using \texttt{screen\_parallel} or one process per GPU using \texttt{screen(ndev=k)}. Figure~\ref{fig:si-multiproc} reports CPU and GPU scaling with prepared conformer libraries.

For CPU scaling, we screened 10{,}000 conformers with \texttt{vol} on an AMD EPYC 9474F node. Throughput increased from 1{,}881 alignments/s with one worker to 102{,}959 alignments/s with 64 workers, giving a 54.7-fold speedup and 85.5\% parallel efficiency (Figure~\ref{fig:si-multiproc}A). Each worker was assigned to a separate physical core and received library entries at regular intervals to balance the work. Worker startup was excluded from timing through an untimed warm-up. The single-worker rate was lower than the 2{,}492 alignments/s measured for screening within the calling process on the same node.

For GPU scaling, we divided 100 million conformers from the expanded prepared pool among one, two, or four NVIDIA L40S GPUs, with one worker process per GPU and one host compute thread per worker (Figure~\ref{fig:si-multiproc}B). Each worker streamed its assigned conformers using the same read-ahead procedure and starting poses as the single-GPU workflow. Screening took 96.69, 48.49, and 24.55\,s on one, two, and four GPUs, respectively, giving 1.99-fold and 3.94-fold speedups relative to one GPU. Screening ten million prepared conformers with \texttt{vol} on one GPU took 9.67\,s.

\clearpage

\begin{table}[H]
\centering
\caption{Software used for the benchmarks. ROSHAMBO2 ran in its own conda environment while all other tools ran in the \sss{} environment. Each column lists the versions installed in that environment. Commits are given for tools built from their repositories. All \sss{} results can be reproduced at commit 0bf8416.}
\label{tab:si-software}
\small
\fitsitable{\begin{tabular}{lll}
\toprule
Software & \sss{} environment & ROSHAMBO2 environment \\
\midrule
\sss{} & 0bf8416 & 0bf8416 \\
ROSHAMBO2 & -- & 1.0.0 (7ff8237) \\
ESP-Sim & 0.0.1 (582b39e) & 0.0.1 (582b39e) \\
Python & 3.10.20 & 3.11.15 \\
RDKit & 2026.03.3 & 2026.03.3 \\
NumPy & 1.26.4 & 2.4.6 \\
SciPy & 1.15.2 & 1.17.1 \\
Numba & 0.59.1 & -- \\
PyTorch & 2.11.0+cu128 & -- \\
Triton & 3.6.0 & -- \\
Open3D & 0.18.0 & -- \\
xtb & 6.7.1 & -- \\
\midrule
\multicolumn{3}{l}{\textit{Standalone executables}} \\
ShaEP & \multicolumn{2}{l}{1.4.0} \\
Align-it & \multicolumn{2}{l}{1.0.4 (fa734f2)} \\
OpenChemLib (PheSA) & \multicolumn{2}{l}{2024.4.2} \\
\bottomrule
\end{tabular}
}
\end{table}

\end{document}